# Simultaneous spectral measurement and programmable filtering on a universal silicon photonic chip

**Yangyang Wan**[1, *], **Rui Tang**[1], **Kazuhiro Ikeda**[2], **Makoto Okano**[2], **Kasidit Toprasertpong**[1], **Mitsuru Takenaka**[1,*]
[1]Department of Electrical Engineering and Information Systems, The University of Tokyo, Tokyo 113-8656, Japan
[2]National Institute of Advanced Industrial Science and Technology, Ibaraki 305-8568, Japan

**Abstract.** Spectral measurement and spectral filtering are two fundamental operations in optics, yet they are conventionally implemented using separate optical systems, hindering their simultaneous and programmable operation on a common platform. Here, we demonstrate a universal silicon photonic chip that unifies spectral measurement and programmable filtering within a single integrated architecture. The chip comprises two 8 × 8 programmable unitary transformation circuits interconnected through optical intensity modulators, forming a universal linear optical circuit. In contrast to conventional universal circuits employing equal-length interconnections, we deliberately introduce different optical path lengths between the two unitary circuits. The resulting wavelength-dependent phase evolution provides spectral diversity while preserving the programmability of the universal circuit, enabling parallel spectral measurement and programmable filtering. The fabricated chip achieves a spectral resolution of 0.05 nm over a bandwidth exceeding 100 nm, simultaneously measures eight independent laser sources, and generates multiple programmable spectral filter functions. We further demonstrate simultaneous spectral measurement and filtering during dynamic reconfiguration of the filter functions. These results establish universal photonic circuits as a versatile platform for integrating spectral sensing and programmable optical processing on a single chip.

***Yangyang Wan, yangyangwan@sjtu.edu.cn, Mitsuru Takenaka, mailto:takenaka@mosfet.t.u-tokyo.ac.jp

## 1 Introduction

Spectral measurement and spectral filtering are two fundamental operations for characterizing and controlling the frequency-domain properties of optical fields.[1,2] Modern optical systems increasingly exploit multiple spectral components to enable information multiplexing, sensing, and computation. For example, optical communication systems use wavelength-division multiplexing to increase transmission capacity;[3,4] spectral imaging techniques acquire hyperspectral information using multi-wavelength illumination;[5,6] and optical computing platforms exploit wavelength channels for parallel processing and task multiplexing. [7,8] In these applications, the spectral content of an optical signal often needs to be measured while selected wavelength components are simultaneously transmitted, suppressed, or otherwise manipulated. Spectral measurement and filtering are therefore complementary functions that are frequently required within the same optical system.[9,10]

Conventionally, spectral measurement and spectral filtering have been implemented as separate optical systems. Representative filtering approaches employ diffraction gratings or prisms,[11,12] multilayer dielectric stacks,[13] optical resonators,[14] and arrayed waveguide gratings.[15] Many of these approaches provide limited programmability after fabrication, making it difficult to realize arbitrary spectral responses. Recent advances in programmable photonic meshes have enabled reconfigurable on-chip spectral filtering, while remaining primarily focused on filtering functionality.[16] Spectrometers, on the other hand, can be broadly categorized into dispersive optics,[17] narrowband filters,[18] Fourier-transform,[19] and reconstructive methods.[20,21] Most conventional spectrometers also lack intrinsic parallelism for simultaneously analyzing multiple independent optical inputs, limiting their scalability in multi-source applications.[22, 23] Although some spectrometers and spectral filters share similar optical components, they are generally designed for distinct functions and do not readily support simultaneous spectral measurement and programmable filtering within the same architecture. Integrating these two functions on a common physical platform could not only reduce system complexity and footprint but also enable closed-loop operation in which spectral information is measured, processed, and used to dynamically reconfigure the optical response. However, simultaneously achieving high-resolution spectral measurement, multi-input parallelism, and programmable spectral filtering within a single integrated architecture remains challenging.

Programmable photonic circuits have recently emerged as versatile platforms for dynamically reconfigurable optical signal processing[24]. In particular, a universal linear optical circuit capable of implementing arbitrary linear transformations can be constructed by cascading two programmable unitary transformation circuits through an array of independently controlled optical amplitude modulators[29]. Each unitary transformation can be implemented using a rectangular mesh of Mach–Zehnder interferometers (MZIs), as introduced by Clements et al.[25, 26], while the intermediate amplitude modulators control the singular values of the transformation. This architecture corresponds to the singular value decomposition of an arbitrary linear transformation and therefore provides a general hardware framework for programmable linear optical processing. In conventional implementations, the two unitary transformation circuits are interconnected through equal-length optical paths to avoid introducing additional wavelength-dependent phase shifts[27]. In contrast, deliberately introducing controlled optical-path-length differences between the two unitary circuits can transform this otherwise unwanted wavelength dependence into an additional degree of freedom, enabling wavelength-selective functionality while retaining the programmability of the universal linear optical circuit[28].

Here, we propose and experimentally demonstrate a programmable spectral measurement and filtering chip (PSMFC) that simultaneously performs reconstructive spectral measurement of multiple optical inputs and programmable spectral filtering. The chip is based on an 8 × 8 universal silicon photonic circuit comprising two programmable unitary transformation circuits implemented with Clements Mach–Zehnder interferometer (MZI) meshes. Unlike conventional universal photonic circuits in which the two unitary circuits are interconnected by equal-length waveguides, we deliberately introduce an array of unequal-length waveguides between them. The resulting wavelength-dependent phase evolution creates spatially diverse spectral responses at the output ports, while the MZI meshes provide programmable control over these responses. The spatially distributed responses are exploited for reconstructive spectral measurement, whereas the programmable wavelength-dependent transformation enables arbitrary spectral filtering, allowing the two functions to operate simultaneously on the same chip. Experimentally, the PSMFC achieves spectral measurement with a spectral resolution of 0.05 nm over a bandwidth exceeding 100 nm, and realizes multiple programmable filter functions. We further demonstrate simultaneous spectral measurement and dynamic filtering in an iterative closed-loop configuration, highlighting the potential of universal photonic circuits as a common platform for integrated spectral sensing and programmable optical processing.

## 2 Results

### *2.1 Concept and principle of PSMFC*

In conventional optical systems, spectral measurement and filtering are typically performed using physically separate components, as illustrated in Fig. 1a. Integrating these two complementary functions into a single programmable photonic circuit could substantially simplify the optical architecture while enabling simultaneous measurement and manipulation of optical spectra. Here, we realize this concept using a PSMFC, which performs spectral measurement and programmable spectral filtering simultaneously within the same photonic circuit (Fig. 1b).

The PSMFC is a multi-input, multi-output silicon photonic chip comprising two Clements-type MZI meshes interconnected through an unequal-length waveguide array. This architecture is based on a universal linear optical processor, in which two programmable unitary transformations are cascaded through a diagonal amplitude-scaling layer to implement arbitrary linear transformations. In conventional universal optical processors, equal optical path lengths are generally maintained between the two unitary meshes to minimize wavelength-dependent phase variations and ensure broadband operation[29]. In the PSMFC, by contrast, controlled path-length differences are deliberately introduced into the intermediate layer to create wavelength-dependent phase shifts, thereby providing an additional spectral degree of freedom (Fig. 1c). The combination of the unequal-length waveguide array and the two programmable unitary meshes produces unique frequency responses at the output ports, generating speckle-like responses distributed across spatial and frequency dimensions. In the spatial domain, the intensity distribution across output ports forms a wavelength-dependent speckle pattern. After calibrating the mapping between wavelength and the corresponding spatial intensity pattern, the input spectrum can be computationally reconstructed from the measured output intensities, enabling reconstructive spectroscopy. For spectral filtering, the frequency response of an individual output port can be programmed by reconfiguring the two MZI meshes, enabling user-defined spectral filter responses. Different output ports can therefore be assigned to different filtering functions, while the remaining ports provide the spatial speckle pattern required for spectral reconstruction. This spatial allocation of the output channels enables spectral measurement and programmable filtering to be performed simultaneously within the same photonic circuit. The finite number of physical output ports would normally limit the number of independent measurement channels available for spectral reconstruction and, consequently, the achievable spectral resolution. To overcome this limitation, additional measurement states can be generated by sequentially reconfiguring the MZI meshes, while constraining the designated filtering ports to preserve their target filter responses. Each configuration produces a distinct spatial spectral fingerprint, effectively increasing the number of measurement channels available to the reconstruction algorithm without requiring additional optical hardware. This reconfigurability therefore decouples the number of effective spectral measurement channels from the number of physical output ports, allowing high-resolution spectral reconstruction and programmable filtering to coexist on a compact photonic chip.

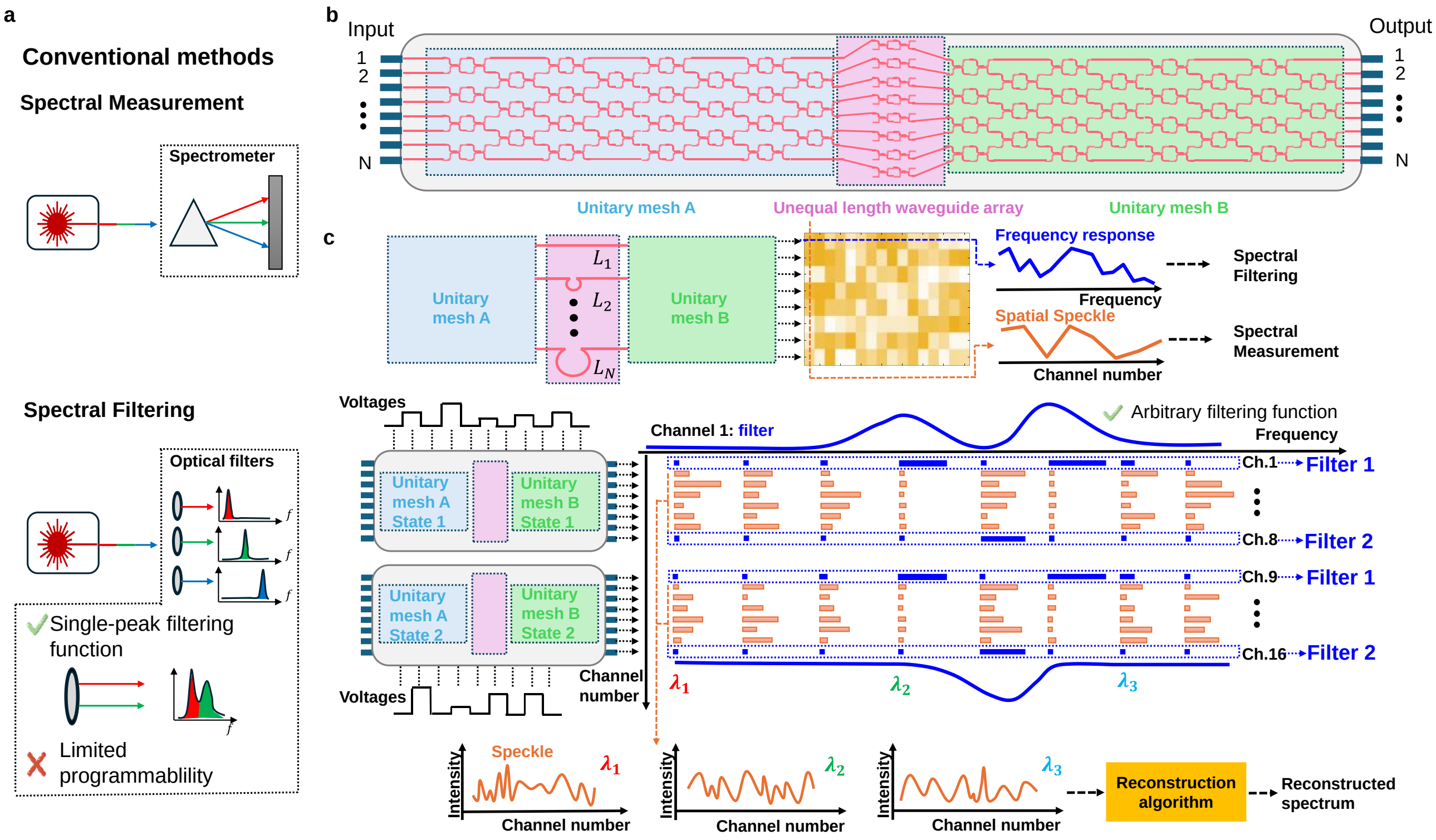

**Fig. 1 Concept and operating principle of programmable spectral measurement and filtering chip (PSMFC). (a)** Conventional spectral measurement and filtering are typically implemented using physically separate systems. Conventional optical filters provide limited programmability, restricting the realization of arbitrary spectral responses. **(b)** Schematic of the proposed PSMFC structure. The chip consists of two programmable unitary transformation circuits (unitary mesh A and unitary mesh B), each implemented using an 8 × 8 Clements MZI mesh, interconnected by an unequal-length waveguide array incorporating amplitude modulators. In contrast to conventional universal linear optical circuits employing equal-length interconnections, the unequal path lengths deliberately introduce wavelength-dependent phase evolution between the two meshes. **(c)** Operating principle of the PSMFC. Due to the presence of an unequal-length waveguide array, PSMFC exhibits speckle response characteristics in both the frequency and spatial dimensions, and this speckle can be programmatically controlled by adjusting mesh A and mesh B. By adjusting PSMFC, the target filters can be obtained at specified output ports, while the spatial speckle at the remaining output ports can be used to reconstruct spectral information via a reconstruction algorithm.

The fabricated PSMFC is shown in Fig. 2. The device has eight input and eight output ports and integrates two Clements-type MZI meshes (Meshes A and B), each comprising 28 MZI cells. The corresponding ports of the two meshes are connected one-to-one through an array of eight waveguides with deliberately different optical path lengths. An additional balanced MZI is incorporated into each interconnecting waveguide to independently control the optical power transferred from mesh A to mesh B, forming a programmable diagonal amplitude-scaling layer between the two unitary meshes. The entire circuit therefore contains 64 MZI cells. Thermo-optic phase shifters are integrated on both arms of every MZI, enabling full reconfigurability of the transmission characteristics through electrical control of the applied voltages. By tuning these phase shifters, the transfer matrix of the mesh A and mesh B can be programmed. The optical path-length distribution of the intermediate waveguide array determines the wavelength-dependent phase evolution between the two meshes and is therefore a key design parameter for both spectral measurement and filtering. Our analysis shows that the maximum and average path-length differences primarily determine the spectral characteristics, whereas the detailed uniformity of the length distribution has a comparatively minor influence (Supplementary Section S1.3). In the fabricated device, we therefore employ a non-uniform distribution of waveguide lengths, with a maximum path-length difference of approximately 420 $\mu m$ and an average path-length difference of approximately 172 $\mu m$.

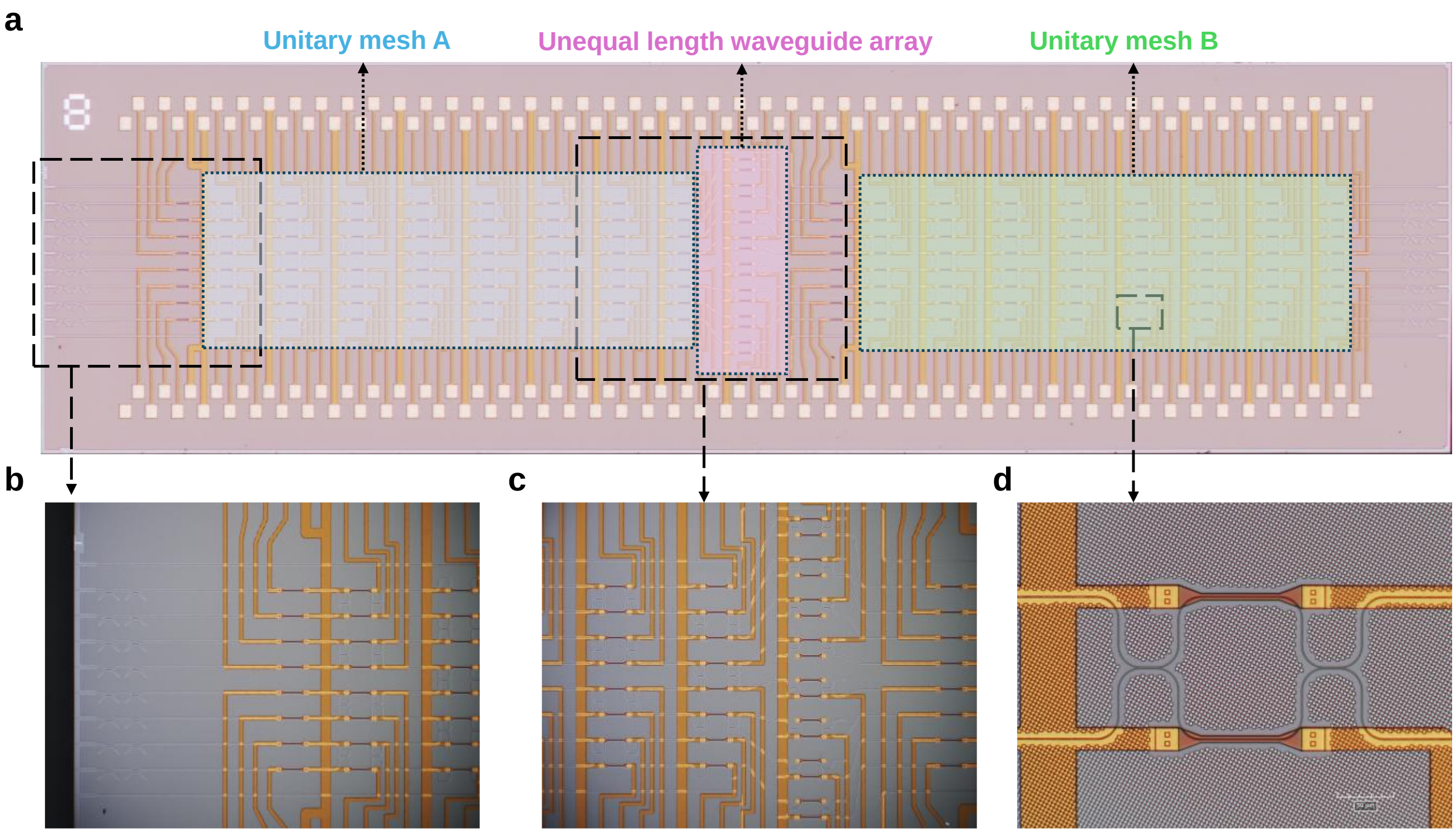

**Fig. 2 Fabricated programmable spectral measurement and filtering chip (PSMFC). (a)** Optical micrograph of the fabricated 8-input, 8-output PSMFC on a silicon photonics platform. The circuit contains 64 MZI cells. **(b)** Enlarged view of the input port region. **(c)** Enlarged view of the unequal-length waveguide array region. A balanced MZI is incorporated into each waveguide for amplitude modulation, providing the additional degree of freedom for programmable control of the circuit. **(d)** Enlarged view of an individual MZI cell.

## *2.2 Spectra measurement*

The operating principle of the PSMFC as a reconstructive spectrometer is illustrated in Fig. 3(a). As with other reconstructive spectrometers, a calibration procedure is first performed to determine the transmission matrix of the system prior to spectral measurement[30, 31, 32].During calibration, a tunable laser is injected into the PSMFC, and the spatial speckles, defined by the intensity distributions across the output ports, are recorded as a function of wavelength. The collection of these wavelength-dependent spatial speckles constitutes the transmission matrix of the spectrometer. To increase the number of effective speckle channels beyond the number of physical output ports, meshes A and B are reconfigured to generate distinct spatial speckles at the same wavelength. Because spectral reconstruction only requires sufficiently distinct speckle responses for different measurement channels, multiple chip configurations can be readily generated using randomly selected control voltages applied to meshes A and B. Figure 3b shows the transmission matrix obtained using eight different chip configurations, while Fig. 3c shows representative spatial speckles at different input wavelengths. Thus, reconfiguration of the PSMFC effectively expands the number of measurement channels without requiring additional physical output ports. Each input port of the PSMFC has its own transmission matrix and can therefore function as an independent reconstructive spectrometer. The calibration procedure is performed separately for each input port, enabling multiple spectrometers to operate in parallel on a single chip. After calibration, multiple lasers under test (LUTs) can be simultaneously injected into different input ports. The spatial speckles are recorded for the different configurations of meshes A and B, and the spectrum at each input port is independently retrieved using the corresponding calibrated transmission matrix and reconstruction algorithm (see Methods). This architecture therefore enables parallel spectral measurement of multiple independent optical inputs using a single PSMFC.

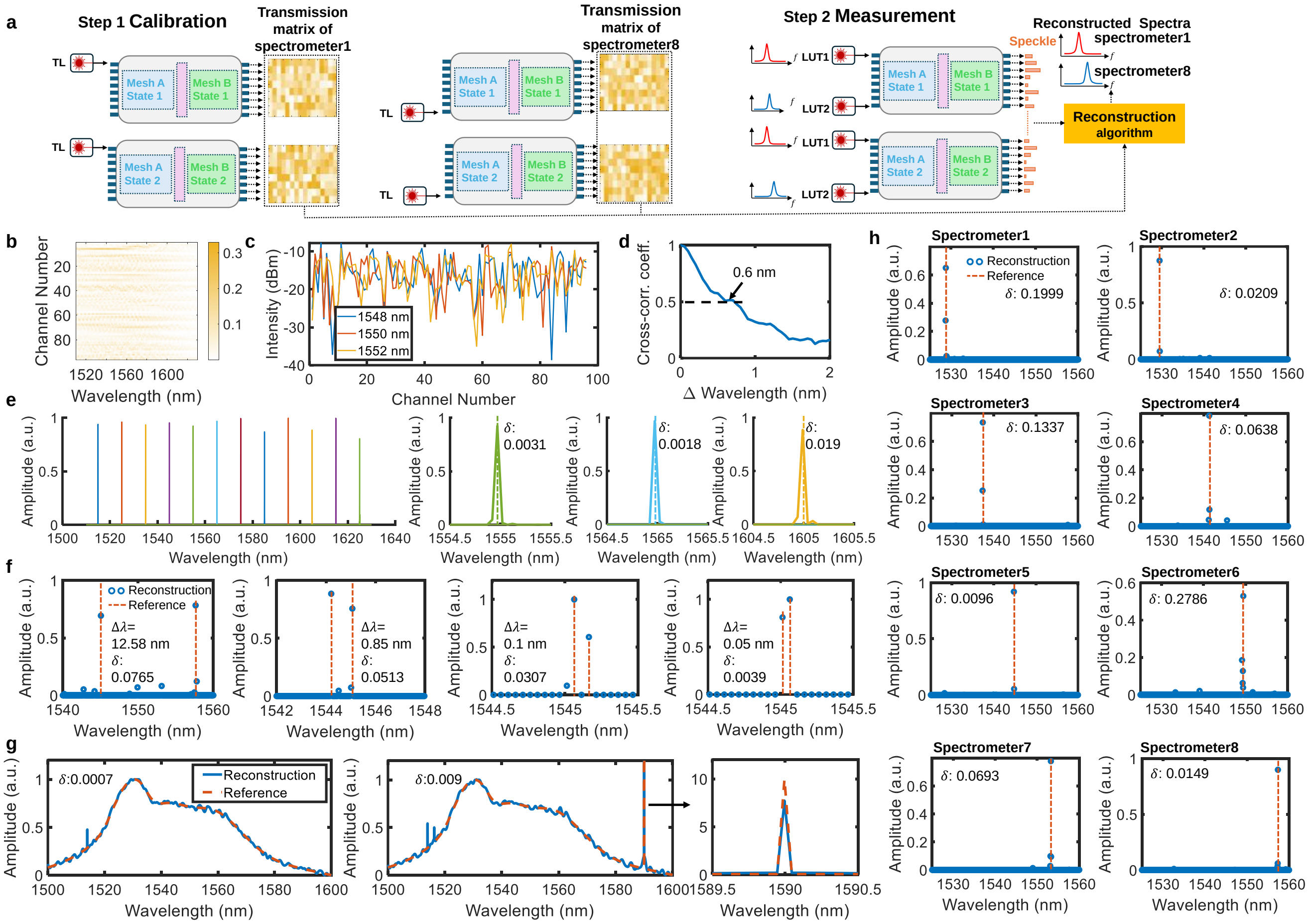

**Fig. 3 Spectral measurement using the PSMFC. (a)** Operation principle of reconstructive spectrometers using the PSMFC. In the calibration step, a tunable laser (TL) is swept across the wavelength range of interest, and the wavelength-dependent spatial speckles at the output ports are recorded to construct a transmission matrix for each input port. Each input port thereby functions as an independent reconstructive spectrometer. Multiple effective measurement channels can be generated by reconfiguring the PSMFC and recording the corresponding spatial speckles. In the measurement step, light sources under test (LUTs) are injected into the input ports, and their spectra are reconstructed from the measured spatial speckles using the calibrated transmission matrices. **(b)** Experimentally measured transmission matrix. **(c)** Spatial speckles measured at representative input wavelengths. **(d)** Cross-correlation

coefficient of spatial speckles as a function of wavelength spacing. The cross-correlation coefficient decreases to 0.5 at a wavelength spacing of approximately 0.6 nm. **(e)** Spectral measurement over a 120-nm wavelength range, demonstrated by reconstructing a series of single-wavelength inputs. **(f)** Reconstruction of dual-wavelength inputs with progressively decreasing wavelength separations, demonstrating the ability to resolve two spectral components separated by 0.05 nm. **(g)** Reconstruction of broadband spectra from an erbium-doped fiber amplifier (EDFA), including measurement of the EDFA spectrum with an additional single-peak component at 1590 nm. **(h)** Simultaneous spectral reconstruction of eight independent laser inputs at different wavelengths.

The presence of unequal length waveguide array allows the PSMFC to be interpreted as a parallel configuration of multiple unbalanced MZIs. Interference among the optical fields from these interferometers generates wavelength-dependent speckle patterns. We developed a physical model of this process and derived its corresponding mathematical formulation (Details in Supplementary Section S1). The analysis reveals that the spectral resolution of the proposed reconstructive spectrometer is fundamentally determined by the maximum optical path difference supported within the system. In reconstructive spectrometers, the half width at half maximum (HWHM) of the cross-correlation function describing the wavelength-dependent evolution of speckle patterns is commonly adopted as a metric to evaluate spectral resolution. In general, a smaller HWHM means the system is more sensitive to wavelength changes and indicates a higher spectral resolution. According to the theoretical model (Details in in Supplementary Section S1), the relationship between the HWHM $\Delta f$ and the maximum waveguide length difference $\Delta L_{\max}$ in the system can be expressed as follows[36]:

$$\Delta f \approx \frac{c}{2n\Delta L_{max}} \tag{1}$$

where $c$ is light speed and $n$ is the effective index of Si waveguide. The speckle cross-correlation curve measured from the fabricated chip is shown in Fig. 3(d). The measured HWHM is 0.6 nm, in close agreement with the theoretical prediction (Details in Supplementary Section S1). To quantify the accuracy of spectral reconstruction, we adopt the widely used L2-norm relative reconstruction error[33, 34], defined as:

$$\varepsilon = \frac{\left\|S_{ref}-S\right\|_2}{\left\|S_{ref}\right\|_2} \tag{2}$$

where $S$ is the reconstructed spectrum and $S_{ref}$ is the reference spectrum. The reconstruction accuracy depends strongly on the number of effective speckle channels. Numerical simulations show that the reconstruction error decreases progressively as the number of channels increases, before approaching saturation beyond a certain channel number (Supplementary Section S1.1). Based on this analysis, we use 12 distinct chip configurations, corresponding to 96 effective speckle channels, for reconstructing sparse spectra. For broadband spectra, which require a larger number of independent measurements for reliable reconstruction, 32 chip configurations are used, providing 256 effective speckle channels.

Fig. 3(e) presents the reconstruction of a series of single-wavelength inputs spanning a spectral range of 120 nm, demonstrating the broad operating bandwidth of the PSMFC. To evaluate the spectral resolution, we measured dual-wavelength inputs with progressively decreasing wavelength separations, from tens of nanometres to tens of picometres (Fig. 3f). Two spectral components separated by as little as 0.05 nm are clearly resolved. We therefore define the experimentally demonstrated spectral resolution as the minimum wavelength separation at which two closely spaced spectral components can be individually resolved, yielding a resolution of 0.05 nm. The system is also capable of measuring broadband spectra as other reconstructive spectrometers. An erbium-doped fiber amplifier (EDFA) source was employed as the LUT, and its broadband spectrum was successfully reconstructed, as shown in Fig. 3g. Accurate reconstruction was also achieved for a composite input consisting of the EDFA output combined with a narrowband laser at 1590 nm, demonstrating the capability to recover both broadband and narrowband spectral features within the same spectrum. Fig. 3(h) demonstrates the parallel measurement of eight laser sources with distinct wavelengths. The small peaks in reconstructed spectra are reconstruction errors, which arise primarily from solving an underdetermined system where the number of spectral channels exceeds that of speckle channels, and additionally from noise, such as temperature variations, that induces speckle deformation.

### *2.3 Spectra filtering*

The operating principle of the PSMFC as a programmable spectral filter is illustrated in Fig. 4(a). It has been theoretically established that a power splitter, an unequal-length waveguide array, and a reconfigurable interferometer mesh can collectively constitute a universal programmable and self-configuring optical filter[28]. In our implementation, mesh A functions as the power splitter, distributing the input optical field into the interconnecting waveguides with different lengths, while mesh B serves as the reconfigurable mesh. For a selected output port designated for filtering, the filter frequency response can be arbitrarily tailored through joint reconfiguration of mesh A and mesh B. To program a target filter response, the spectral transmission at the designated output port is first measured by sweeping a tunable laser, while the control voltages applied to the PSMFC are iteratively optimized to minimize the deviation between the measured and target responses. A hybrid optimization algorithm combining a genetic algorithm (GA) and simulated annealing (SA) is employed to efficiently search the large parameter space and converge toward the desired response. Once the optimization is completed, the resulting voltage configuration is stored and can subsequently be applied to reproduce the programmed filter response. Fig. 4(b) shows representative spectral filter responses programmed at the same designated output port using different voltage configurations. For filter response 1, a single-passband response centred at 1550.75 nm is programmed at output port 4. The measured response closely follows the target profile, with the optical power at the passband centre predominantly directed to the designated output port, as shown by the corresponding output power distribution. By reprogramming the same PSMFC without modifying the physical circuit, substantially different spectral responses can be obtained at the same output port. Filter responses 2 and 3 demonstrate more complex user-defined spectral functions containing multiple transmission peaks and valleys, for which the measured responses also closely reproduce the target profiles. These results demonstrate that the spectral response is not fixed by the physical geometry of the unequal-length waveguide array but can be dynamically synthesized through coherent control of the optical fields by the two programmable MZI meshes. The same physical output port can therefore be reconfigured to implement different spectral filter functions simply by changing the electrical control voltages. The multi-output architecture further allows different spectral filter functions to be programmed simultaneously at different output ports. Fig. 4(c) demonstrates two single-passband filters implemented at separate output ports, with centre wavelengths of 1548.5 nm and 1551 nm, respectively. When the input wavelength is tuned to the centre wavelength of either filter, the optical power is predominantly directed to its designated output port, as evidenced by the corresponding output power distributions in Fig. 4c. These results demonstrate that the PSMFC can simultaneously implement multiple independently programmable spectral filter functions within a single photonic circuit.

The filtering characteristics of the PSMFC were further analyzed using a theoretical model (Supplementary Section S1.2). The minimum achievable passband width $\Delta f_{filter}$ is fundamentally determined by the maximum optical path difference in the system ($\Delta L_{max}$), whereas the overall operational bandwidth $f_{FSR-eff}$ of the filter is primarily governed by the average optical path difference ($\Delta L_{\text{ave}}$). In the frequency domain, these quantities scale approximately as $\Delta f_{filter} = c/n\Delta L_{max}$ and $f_{FSR-eff} = c/n\Delta L_{\text{ave}}$, respectively, where $c$ is the speed of light and $n$ is the effective index. For the fabricated PSMFC, $\Delta f_{filter}$ and $f_{FSR-eff}$ are approximately 1.2 nm and 4.5 nm, respectively, in good agreement with the experimentally observed filter characteristics.

As shown in Fig. 4(b) and 4(c), small deviations remain between the experimentally realized and target filter responses. Numerical analysis indicates that this error decreases with increasing chip scale (Details in Supplementary S1.2). In the present device, only eight unequal length waveguides are implemented, which inevitably limits the achievable accuracy. Further improvement in filtering precision can be expected with larger-scale implementations incorporating a greater number of path-length variations into the unbalanced waveguide array.

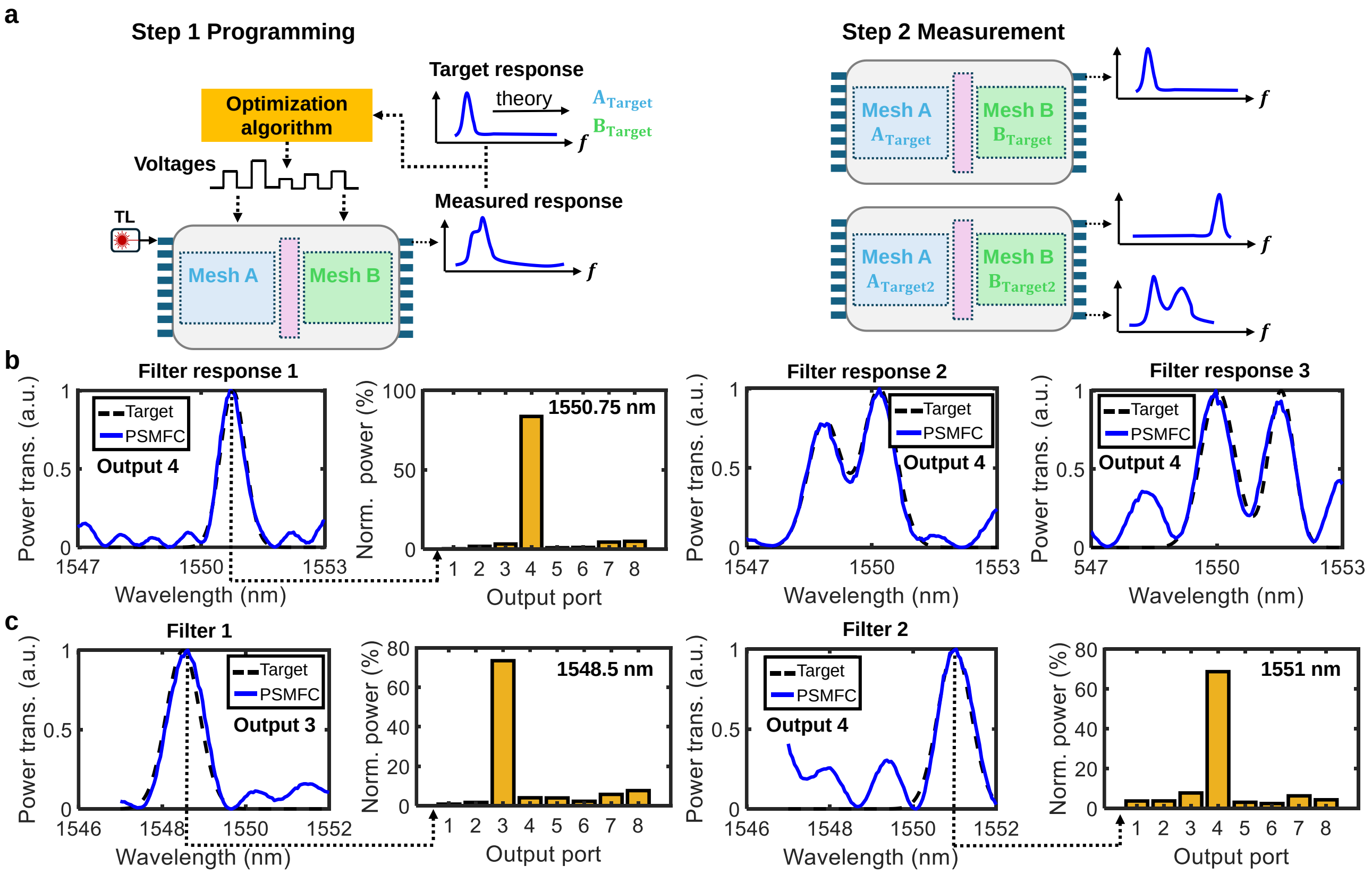

**Fig. 4 Programmable spectral filtering using the PSMFC. (a)** Operating principle of programmable spectral filtering. In the programming step, the spectral response at a designated output port is measured by sweeping a tunable laser (TL), and an optimization algorithm iteratively adjusts the control voltages applied to meshes A and B to minimize the deviation from the target filter response. Once optimized, the control voltages corresponding to each target response are stored and subsequently applied to configure the desired filter function. **(b)** Different spectral filter functions programmed at the same output port. Filter response 1 is a single-peak bandpass filter centered at 1550.75 nm; the corresponding optical power distribution across the eight output ports at the center wavelength is also shown. Filter

responses 2 and 3 demonstrate user-defined spectral responses obtained by applying different sets of control voltages **(c)** Simultaneous implementation of two different spectral filters at separate output ports. The optical power distributions across the eight output ports are shown at the center wavelength of each bandpass filter.

### *2.4 Simultaneous spectral filtering and spectral measurement*

Spectral measurement and programmable filtering can be performed simultaneously by assigning different output ports of the PSMFC to the two functionalities, as illustrated in Fig. 5(a). During the initial programming and calibration step, an optimization algorithm is used to configure a designated output port such that its spectral response matches the target filter function, while the spatial speckles at the remaining output ports are recorded to construct the transmission matrix for spectral reconstruction. Because the current PSMFC has only eight physical output ports, multiple chip configurations are employed to increase the number of effective speckle channels available for reconstruction, as exemplified by State 1 and State 2 in Fig. 5(a). For each state, the same target constraint is imposed on the designated filtering port during programming. Because only a subset of the output ports is constrained by the target filter function, the remaining degrees of freedom of the programmable circuit can be exploited to identify multiple voltage configurations that produce consistent, though not exactly identical, filter responses at the designated port but distinct spatial speckles at the other ports. This behavior is experimentally demonstrated in Fig. 5(b) and 5(c): the spectral response of the filtering port remains nearly unchanged between State 1 and State 2 (Fig. 5(b)), whereas the optical power distributions at the remaining output ports differ substantially between the two states (Fig. 5(c)). These distinct speckles provide additional effective measurement channels without compromising the programmed filtering function. After calibration, the LUT is injected into the chip. The selected port performs spectral filtering, while the speckle patterns at the remaining ports enable reconstruction of the incident spectrum. Fig. 5(d) shows the spectral reconstruction results for two different incident wavelengths when output port 4 is configured as a bandpass filter, with the corresponding filter response shown in Fig. 5(b). The two cases correspond to an incident wavelength within the passband (1550 nm) and a wavelength suppressed at the filtering port (1549 nm), respectively. In both cases, the incident wavelength can be successfully reconstructed from the speckle responses at the remaining output ports, demonstrating that spectral measurement remains available regardless of whether the incident wavelength is transmitted or suppressed by the filtering port.

In practical applications, dynamic filter reconfiguration and subsequent spectral measurement feedback are often required within a closed-loop process to implement adaptive filtering and monitor its performance. Conventional approaches typically require separate spectral measurement and filtering modules. In contrast, the proposed scheme integrates these functions within a single PSMFC, as illustrated in Fig. 5(e). Before operation, the PSMFC is programmed for a set of target filter responses, and the corresponding control-voltage configurations and transmission matrices for spectral reconstruction are obtained and stored. The operating procedure consists of three steps. First, the incident light under test (LUT) is injected into the PSMFC, which operates as a spectrometer to reconstruct the input spectrum from the spatial speckle responses. Second, according to the measured spectrum and the desired filtering function, the corresponding voltage configuration is applied to program the designated output port as the target spectral filter. Third, the filtered light from the designated output port is routed back to one of the remaining input ports of the PSMFC, allowing the same chip to reconstruct the spectrum after filtering. To increase the number of effective speckle channels, the filtered output can be sequentially routed to different available input ports. With seven available input ports and seven output ports, a total of 49 effective speckle channels can be generated. Because the spectral filter operates over a finite bandwidth determined by the optical path-length distribution of the PSMFC, reconstruction of the filtered spectrum can be restricted to the relevant wavelength range of 1547–1553 nm. With a spectral resolution of 0.05 nm, reconstruction of the resulting 120-point spectrum from 49 speckle channels is feasible.

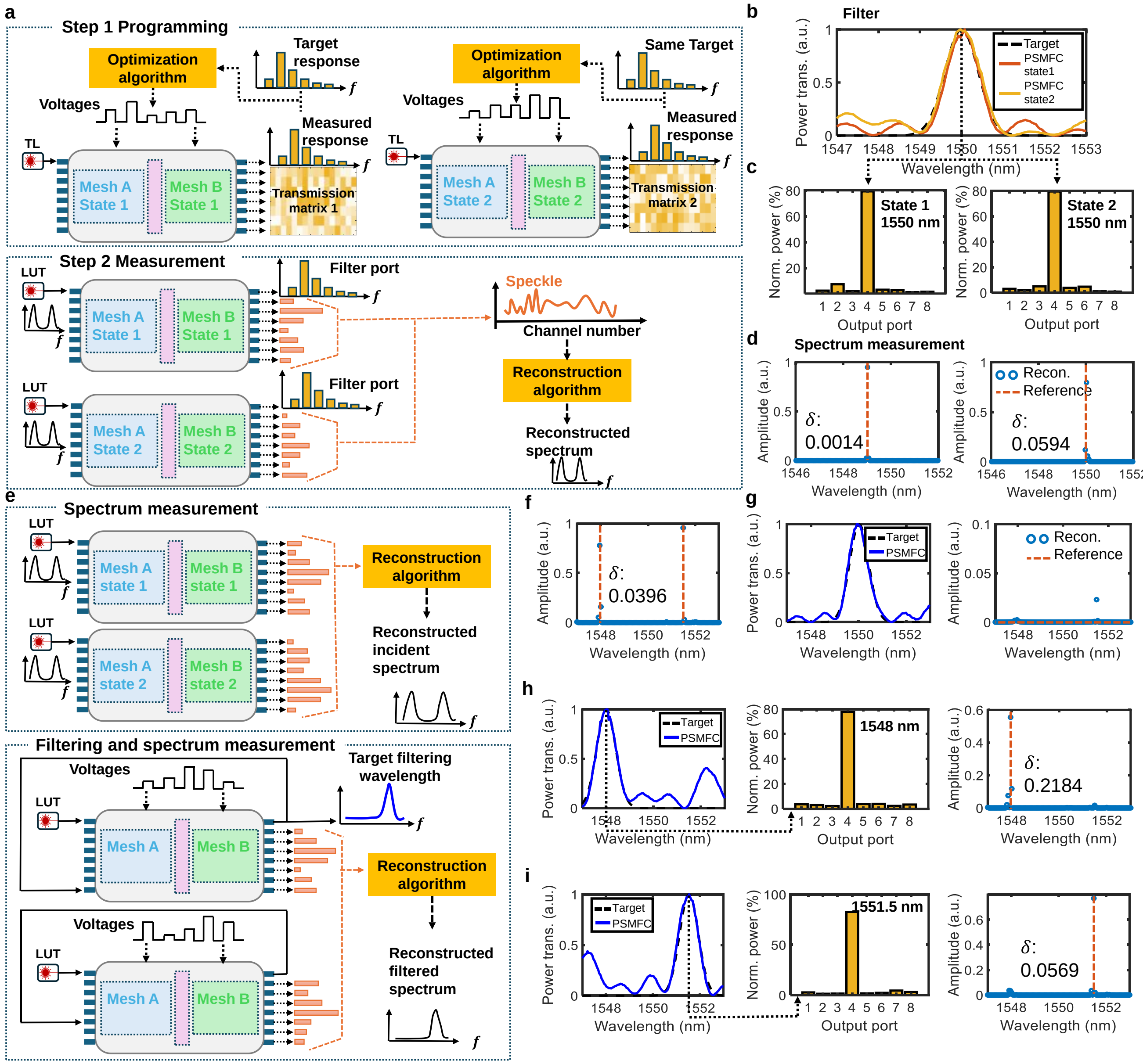

**Fig. 5 Simultaneous spectral filtering and measurement using the PSMFC. (a)** Working principle. During calibration, the control voltage corresponding to the target filtering response and the transmission matrices of other output ports are obtained. To increase spatial speckle channels, the PSMFC state is varied while maintaining a nearly unchanged filter response. After calibration, the light under test (LUT) is injected. By tuning the chip state, the filter port performs the desired filtering while spatial speckle is collected for spectral reconstruction. **(b)** The filter port maintains a consistent response under different PSMFC states. **(c)** Output intensity distributions under different chip states when the LUT wavelength equals the bandpass filter's center wavelength. The filtering function remains unchanged, while leakage distributions at other ports vary. **(d)** Reconstructed spectra when the LUT wavelengths are 1549 nm and 1550 nm while implementing the filter in (b). **(e)** Dynamic filter reconfiguration using a single PSMFC. The spectrometer function first measures the LUT spectrum, then the control voltage is adjusted to realize the desired filter response. The filtered spectrum is monitored by connecting the filter port to other input ports sequentially to increase speckle channels. Dynamic filtering can be achieved by changing control voltages. **(f)** Reconstructed spectrum of a dual-peak LUT at 1548 nm and 1551.5 nm. **(g)** Frequency response of the PSMFC filter and reconstructed spectrum of the filtered light. The initial filter is a 1550 nm-centered bandpass, under which both peaks are suppressed. **(h)** Reconfigured bandpass filter centered at 1548 nm, showing its response, output power distribution, and reconstructed filtered spectrum of the LUT in (f). **(i)** Bandpass filter reconfigured to 1551.5 nm, with the reconstructed spectrum of the filtered LUT in (f).

The experimental results corresponding to this operating procedure are presented in Figs. 5(f)–5(i). Fig. 5(f) shows the reconstructed spectrum of the incident LUT, which exhibits two spectral components at 1548 nm and 1551.5 nm. The filter is then initially configured as a bandpass centered at 1550 nm, as shown in Fig. 5(g). Because the passband lies between the two spectral components of the LUT, both components are strongly suppressed in the filtered output, which is confirmed by the reconstructed spectrum after filtering. Fig. 5(h) and 5(i) further demonstrate dynamic reconfiguration of the spectral filter by changing the electrical control voltages. When the passband is shifted to 1548 nm (Fig. 5(h)), the 1548 nm component is selectively transmitted while the 1551.5 nm component is suppressed. Conversely, when the filter is reconfigured with a passband centered at 1551.5 nm (Fig. 5(i)), the 1551.5 nm component is selectively transmitted while the 1548 nm component is suppressed. The corresponding output power distributions at the passband center wavelengths confirm that the transmitted light is predominantly directed to the designated filtering port, while the reconstructed spectra of the filtered outputs further verify the intended wavelength-selective operation. These results demonstrate that the PSMFC can dynamically reconfigure its spectral filtering function while providing spectral feedback of the filtered signal within the same programmable photonic platform.

## 3 Discussion

Long-term stability is an important consideration for both programmable spectral filters and reconstructive spectrometers[35, 36].In the PSMFC, temperature variations represent a major source of instability because of the strong thermo-optic response of silicon. A spatially uniform temperature variation primarily produces a systematic spectral shift of the filter response and a corresponding wavelength offset in the reconstructed spectrum. For example, an input wavelength of 1550 nm may be reconstructed at a shifted wavelength under temperature variation. Therefore, temperature drift mainly affects the wavelength accuracy of the spectral measurement rather than its precision[37]. In typical silicon photonic circuits, a temperature variation of 1°C corresponds to an effective spectral shift of approximately 0.08 nm[38, 39]. Spatially nonuniform temperature variations are potentially more detrimental, as they alter the relative phases among different optical paths and can therefore distort both the programmed filter response and the calibrated spatial speckles used for spectral reconstruction. Thermal stabilization is therefore particularly important for high-resolution spectral reconstruction. In our system, the spectrometer resolution reaches 0.05 nm, imposing higher temperature control requirement than filter function. Temperature fluctuations originate from both environmental variations and internal thermal effects. Since the chip integrates 64 MZIs equipped with thermo-optic phase shifters, the reconfiguration of the PSMFC will change the heat distribution on the chip, which in turn affects thermal stability and the performance of spectral measurement and filtering. To mitigate temperature variations, we implemented a real-time temperature-feedback system to stabilize the chip temperature and evaluated the long-term stability of spectral reconstruction over 10 h (Details in Supplementary Section S3). The measurements confirm that, under stabilized thermal conditions, accurate spectrum reconstruction can be maintained over long term. Further improvements could be achieved by employing more advanced thermal stabilization techniques[40, 41] or by replacing thermo-optic tuning with alternative phase-shifting mechanisms, such as electro-optic modulation[42, 43, 44], thereby reducing self-heating and thermal crosstalk in future large-scale implementations.

The key performance metrics of the PSMFC are fundamentally determined by the unequal-length waveguide array. Larger optical path differences and a greater number of waveguides enable higher-resolution spectral reconstruction and finer filter functions. However, increasing these parameters inevitably enlarges the chip footprint and introduces additional fabrication complexity. For filtering applications, optical loss must also be carefully considered. The current PSMFC employs edge coupling, and the total loss combined coupling and propagation loss for each input port is approximately 10 dB (Supplementary materials S4.1). Scaling the chip to larger dimensions would further increase propagation loss, which is detrimental to the realization of low-loss filters. Future improvements may be achieved by adopting ultra-low-loss silicon waveguide technologies or by implementing the architecture on alternative low-loss platforms, such as silicon nitride[45, 46, 47].

In conclusion, we have demonstrated a programmable silicon photonic architecture that unifies spectral measurement and filtering within a single integrated platform. By deliberately introducing unequal optical path lengths between two programmable MZI meshes, wavelength dependence, which is typically minimized in universal linear optical circuits, is harnessed as an additional degree of freedom for spectral processing. The resulting wavelength-dependent spatial speckles enable reconstructive spectral measurement with a demonstrated resolution of 0.05 nm over a bandwidth exceeding 100 nm, together with parallel measurement of eight independent optical inputs. At the same time, arbitrary filter functions are obtained for spectral filtering through programmable control of the output port frequency responses. Importantly, the same PSMFC can perform spectral measurement and programmable filtering simultaneously by assigning different output ports to the two functions. We further demonstrated dynamic filter reconfiguration with spectral feedback of the filtered output using the same chip, illustrating how spectral measurement and manipulation can be combined within a unified programmable photonic system. More broadly, our approach transforms engineered wavelength dependence from a limitation of universal photonic circuits into a functional resource, providing a route toward multifunctional photonic processors that integrate spectral sensing, manipulation, and reconfiguration on a common hardware platform.

## 4 Methods

### Device fabrication

The chip was fabricated by National Institute of Advanced Industrial Science and Technology (AIST) using ArF immersion lithography on a silicon-based platform. The width of the silicon waveguide is 440 nm. The chip contains a total of 64 MZIs. Each MZI is a balanced MZI with a 50:50 splitting ratio and features thermal phase shifters on both its upper and lower arms. The MZI arm length is 75 $\mu m$. The overall chip size is 2.94 mm $\times$ 9.5 mm, and edge coupling is employed.

### Experiments

In the experiments, the laser source is routed through a 1 × 8 optical switch (Keysight, N7734A),

enabling selective coupling of the input light into any of the eight input ports of the chip. The eight outputs of the switch are connected, via polarization controllers, to the corresponding input ports of the PSMFC. The eight output ports of the chip are connected to a multichannel optical power meter (Santec, OPM-150) for power monitoring. The thermo-optic phase shifters integrated on the chip are driven by a multichannel voltage controller (Aisthesis Co., Ltd., 200-Channel DAC Array). The laser source, optical switch, multichannel optical power meter, and multichannel voltage controller are all remotely controlled via a computer using a Python-based control platform, enabling wavelength tuning, channel selection, optical power acquisition, and voltage programming. During calibration, a single tunable laser source (Santec, TSL-570) is typically employed. In the measurement stage, dual-wavelength spectra are generated by combining two tunable lasers through a fiber coupler; broadband spectra are provided by an erbium-doped fiber amplifier (EDFA, Luxpert, LXI 2000); broadband spectra with superimposed narrowband peaks are realized by combining the EDFA and a tunable laser via a fiber coupler; and multi-source measurements are conducted using an eight-channel laser array (Think-Lands Co., Ltd, Multi-Channel Tunable Laser).

Active temperature stabilization is implemented for the PSMFC. A thermistor sensor (Thorlabs, TH10K) is attached near the chip to monitor its temperature. A thermoelectric cooler is mounted beneath the chip substrate to regulate its temperature. The temperature information acquired by the sensor is fed back to a temperature controller (Thorlabs, TED200C), which applies a proportional–integral–derivative (PID) algorithm to determine the appropriate control current supplied to the TEC, thereby maintaining thermal stability via closed-loop feedback. In the experiments, the chip temperature is stabilized at 40 °C.

**Reconstruction algorithm**

The reconstruction algorithm used for spectral information reconstruction is an optimization algorithm based on convex optimization (CVX)[48], and its expression is as follows:

$$Minimize(\left\| I - \sum_{i=1}^{N} T_i S_i \right\|_2 + \zeta \sum_{i=1}^{N} \|S_i\|_1 + \varsigma \sum_{i=1}^{N} \|D(S_i)\|_2) \tag{3}$$

where $I$ is the measured speckle, $S_i$ and $T_i$ are the reconstructed spectral result and transfer matrix of the i-th spectrometer, respectively. $N$ is the number of spectrometers. $D$ is the differential operation function. $\zeta$ and $\varsigma$ are the coefficients of the L1-norm and L2-norm regularization terms, respectively. The L1-norm and L2-norm regularization terms respectively constrain the sparsity and smoothness of the reconstructed spectrum.

**Optimization algorithm**

The optimization algorithm is employed to determine the control voltages of the PSMFC in order to realize a target spectral response for filtering. According to the theoretical model, a desired frequency response uniquely specifies the corresponding states of mesh A and mesh B, from which the ideal control voltages can be derived under perfect conditions. In practice, however, fabrication imperfections and thermal crosstalk, particularly in a densely integrated chip comprising 64 MZIs, cause deviations from the ideal model, such that the theoretically predicted voltages often fail to produce the intended filter response. An optimization procedure is therefore introduced to compensate for these non-idealities and accurately program the target filter function.

The optimization framework combines a genetic algorithm (GA) with simulated annealing (SA)[49, 50, 51]. Starting from the theoretically estimated control voltages, random perturbations are introduced to generate the initial GA population. After every five GA iterations, the best-performing individual is further refined using the SA algorithm to escape local minima and achieve improved convergence. Both GA and SA employ the same loss function, defined as follows:

$$Minimize(\sum_{m=1}^{N}\left\|F_{mea_m}(f)-F_{tar_m}(f)\right\|_2+\sum_j W_j\sum_{i=1}^{8}\left\|P_{mea_m_j}(i)-P_{tar_m_j}(i)\right\|_2) \quad (4)$$

where $F_{mea_m}$ and $F_{tar_m}$ represent the measured and target frequency responses of the m-th filter, respectively. There are $N$ filters in total. $P_{mea_m_j}(i)$ and $P_{tar_m_j}(i)$ represent the measured power and target power of the i-th output port of the chip at the target filtering peak wavelength, respectively, for the m-th filter. $W$ is a weighting coefficient. $j$ is the number of peak responses in the target filtering function. Convergence is typically achieved within 100 epochs.

**Acknowledgements** This work was partly supported by JST, CREST (JPMJCR2004 and JPMJCR26R3 received by M.T.), JST, MIRAI (JPMJMI20A1 received by M.T.).

**Author contributions** Y.W. contributed to ideas, measurement, and manuscript preparation, R.T contributed to design and measurement, K.I. and M.O. contributed to design and chip fabrication, K.T. contributed to the overall discussion, M.T. contributed to ideas, discussion, and manuscript preparation.

**Data availability** The data that support the findings of this study are available from the corresponding author upon reasonable request.

**Code Availability** The code used in this study is available from the corresponding author upon request.

**Author contributions** Y.W. conceived the idea, performed the experiments, analyzed the results and prepared the manuscript. R.T. designed the chip. K.T. and M.T. discussed the work. M.T. supervised the project and revised the paper.

**Conflict of interest** The authors declare no competing interests.

# Simultaneous spectral measurement and programmable filtering on a universal silicon photonic chip

**Yangyang Wan[1,*], Rui Tang[1], Kazuhiro Ikeda[2], Makoto Okano[2], Kasidit Toprasertpong[1], Mitsuru Takenaka[1,*]**

[1]Department of Electrical Engineering and Information Systems, The University of Tokyo, Tokyo 113-8656, Japan

[2]National Institute of Advanced Industrial Science and Technology, Ibaraki 305-8568, Japan

* *Corresponding author: yangyangwan@sjtu.edu.cn, takenaka@mosfet.t.u-tokyo.ac.jp*

## Contents

## S1 Theoretical analysis for spectral measurement and filtering

### S1.1 Theoretical analysis for spectral measurement

#### S1.1.1 Generation of spatial speckle

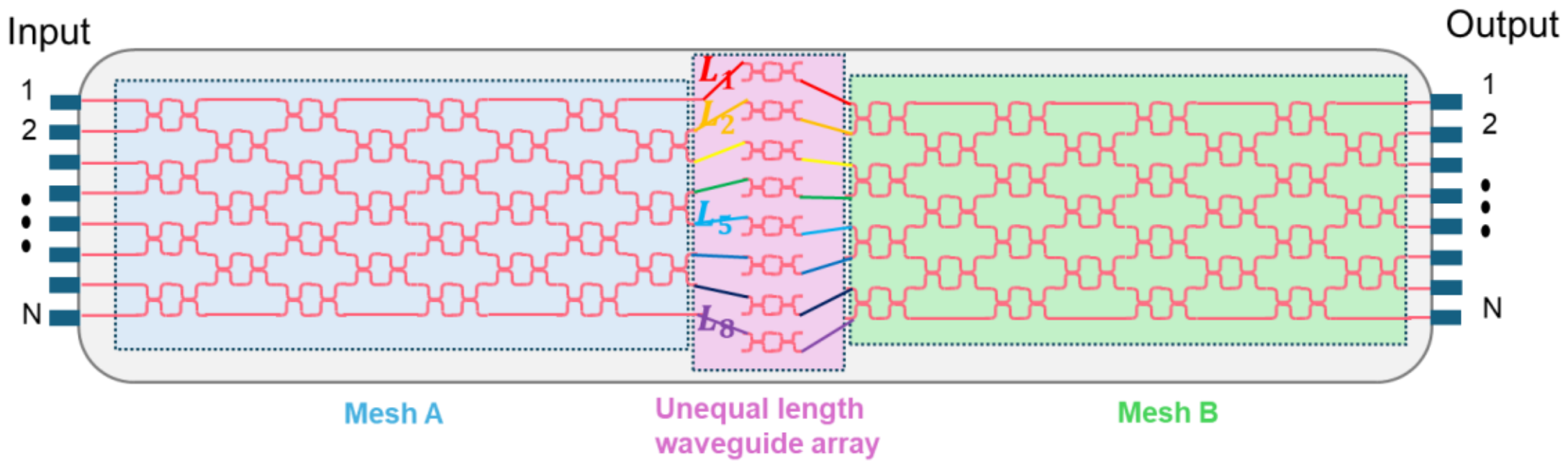

**Fig. S1. Structure of the programmable photonic spectral measurement and filtering chip (PSMFC).**

The structure of the PSMFC is shown in Fig. S1. When the input light is injected into the q-th input port, the optical field emerging from the j-th port of Mesh A and propagating through the corresponding waveguide with length $L_j$ can be expressed as:

$$A_{qj}C_j\exp\left(-iknL_j\right) \tag{1}$$

where $A_{qj}$ and $C_j$ are the coefficients of Mesh A and the unequal-length waveguide array, respectively. At the output port $p$ of the PSMFC, the optical field can be written as:

$$E_{out} = \sum_{j=1}^{m} B_{jp}A_{qj}C_j\exp\left(-iknL_j\right) \tag{2}$$

where $B_{jp}$ is the coefficient of Mesh B for the transmission from input port $j$ to output port $p$, and $m$ is the total number of waveguides in the waveguide array. Accordingly, the optical intensity at port $p$ can be obtained as:

$$I_p \propto 2\sum_{j=1}^{m}\sum_{i=1}^{m} B_{jp}A_{qj}C_jB_{ip}A_{qi}C_i\cos\left(\frac{2\pi n\Delta L_{ij}}{\lambda}\right) \tag{3}$$

where $\Delta L_{ij}$ is the length difference between the i-th and j-th waveguides. From Eq. (3), it can be seen that different output ports $p$ produce different intensity values due to the distinct coefficients of Mesh A and Mesh B, thereby forming spatial speckle patterns. Because the coefficients of Mesh A and Mesh B are tunable, the number of spatial speckle channels can be further expanded by reconfiguring these meshes. In other words, different speckle intensity patterns can be obtained at the same output port by applying different configurations of Mesh A and Mesh B. Moreover, when the wavelength of the input light varies, the spatial speckle pattern changes accordingly. This property enables the implementation of a reconstructive spectrometer.

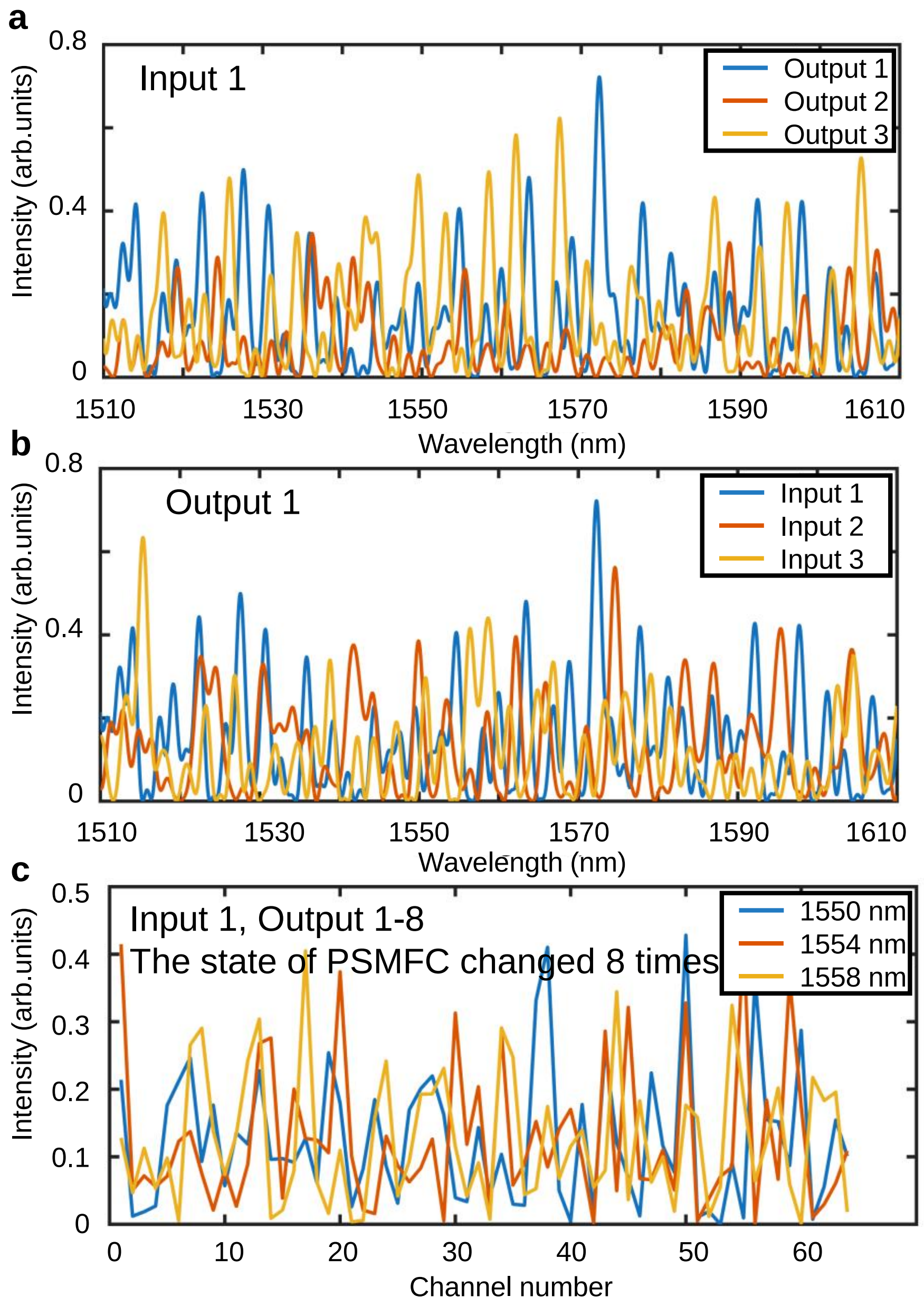

**Fig. S2. Simulated speckle results.**

We performed numerical simulations based on the geometrical dimensions and structural parameters of the fabricated chip. The results are shown in Fig. S2. Fig. S2(a) presents the spectral responses of the three output ports when the input light is injected into Input 1. It can be observed that the spectral responses differ among the output ports because each port corresponds to different coefficients in Mesh A and Mesh B. Fig. S2(b) shows that different spectral responses can also be obtained at the same output port (Port 1) when the input port is changed. Therefore, for a fixed input, the spatial speckle pattern at the chip output varies with wavelength, as illustrated in Fig. S2(c). Furthermore, by reconfiguring the state of the PSMFC, the number of spatial speckle channels can be further increased.

### S1.1.2 Theoretical resolution

Intuitively, the lower the similarity between the spatial speckle patterns corresponding to adjacent wavelengths, the easier it is to distinguish them, leading to a higher spectral resolution. The similarity between speckle patterns at different wavelengths is typically quantified by calculating their correlation coefficient. The cross-correlation is defined as follows (For a detailed derivation, please refer to please refer to reference: Wan, Y., Zhou, Q., Ma, L., Fan, X., & He, Z. (2026). Resonant microtaper leaky-mode computational spectropolarimetry with tens of femtometers spectral resolution and full Stokes measurement. Nature Communications, 17, 6341.):

$$C(\Delta k) = \frac{\langle I(k_1)I(k_2)\rangle}{\sqrt{\langle I(k_1)\rangle^2 \langle I(k_2)\rangle^2}} \tag{4}$$

where $k$ is the propagation constant, $I(k_1)$ is the spatial speckle pattern with a propagation constant of $k_1$. Equation (4) can be further simplified as:

$$C(\Delta k) = \frac{\sin{(\Delta k n \Delta L_{max})}^2}{(\Delta k n \Delta L_{max})^2} \tag{5}$$

From this result, it can be seen that the correlation coefficient becomes zero when the condition

$$\Delta k_0 n \Delta L_{max} = \pi \tag{6}$$

is satisfied. The relationship between optical frequency variation and propagation constant difference is

$$\Delta k_0 = \frac{2\pi \Delta \mathrm{f}_0}{\mathrm{c}} \tag{7}$$

By substituting Eq. (6) into Eq. (6), we can obtain

$$\Delta f_0 \approx \frac{c}{2n\Delta L_{max}} \tag{8}$$

When the optical frequency variation reaches the value of Eq. (8), the two speckles are no longer correlated, and the distinction between speckles can be achieved. Therefore, $\Delta f_0$ can be approximated as the system's resolution, and this resolution is determined the maximum optical path difference. Accordingly, a smaller $\Delta f_0$ indicates a higher spectral resolution. Wavelength spacing can be calculated based on the relationship between optical frequency and wavelength. Alternatively, optical frequency variation can be represented by wavelength spacing based on the conversion relationship between wavelength and optical frequency, as shown in Fig. S3.

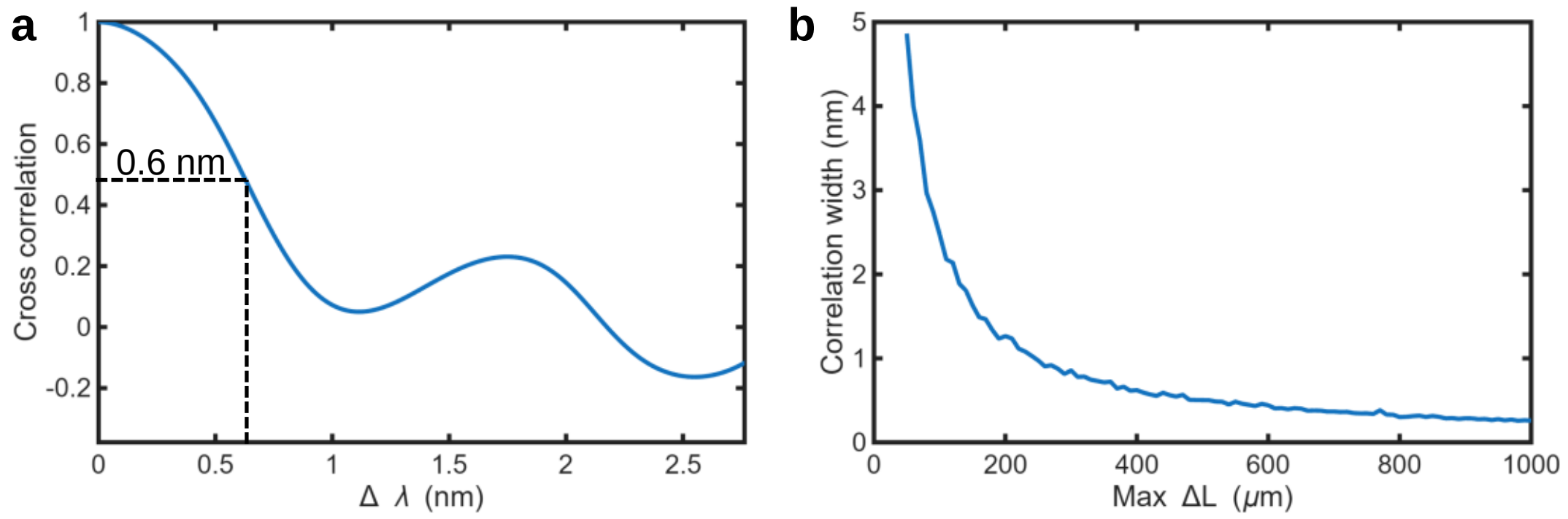

**Fig. S3. Simulation results. (a) Cross-correlation coefficient versus wavelength spacing at a maximum length difference of 420 µm. (b) Correlation coefficient half-width at half-maximum (HWHM) versus maximum length difference between waveguides in the system.**

We performed numerical simulations to analyze the system performance. In the simulations, the number of unequal-length waveguides was fixed at eight (See the design of the length of the array waveguides in Supplementary Section S4.1). A maximum length difference between the waveguides was specified, and eight waveguide lengths were then generated with a uniform random distribution within this range. The wavelength of the incident light was subsequently varied, and the cross-correlation coefficients between the speckle patterns corresponding to different wavelengths were calculated. The results are shown in Fig. S3. Fig. S3(a) shows the cross-correlation curve obtained for a maximum length difference of 420 µm. The corresponding correlation width, defined as the wavelength spacing at which the normalized cross-correlation coefficient between two speckles decreases to 0.5, is approximately 0.6 nm. This value is usually considered as the spectral resolution, and agrees well with the experimental result shown in Fig. 3(d) of the main text. Figure S3(b) shows the correlation width as a function of the maximum length difference. The results are consistent with the theoretical prediction of Eq (6), indicating that a larger maximum optical path difference leads to a higher spectral resolution of the system.

### S1.1.3 Reconstruction principle

Before performing spectral measurements with a reconstructive spectrometer, a calibration procedure is required to obtain the system transmission matrix. This transmission matrix $T$ is composed of the speckle patterns corresponding to different incident wavelengths. After the transmission matrix is established, the measured speckle pattern $I$ and the spectrum $S$ of the unknown input light satisfy the following relationship:

$$S_N = T_{M\times N}^{-1} I_M \tag{7}$$

Where, $M$ is the number of speckle channels, and $N$ is the number of spectral channels to be reconstructed. When $N > M$, the Eq (7) becomes an ill-posed problem, making it difficult to guarantee a unique and accurate solution. Therefore, many studies tried to improve spectral reconstruction based on the transmission-matrix framework by employing optimization algorithms or compressed sensing techniques to achieve better reconstruction performance.

### S1.1.4 Reconstruction error

The reconstruction error arises from two main sources. One is the variation of speckle patterns caused by noise, and the other is the number of speckle channels used for reconstruction. In the presence of noise, the measured speckle pattern deviates from the speckle pattern obtained during calibration, thereby introducing reconstruction errors. Therefore, noise sources in the system should be minimized as much as possible.

Here, we mainly focus on the second factor, namely the speckle channel number. As analyzed in Section S1.1.3, when the number of speckle channels is smaller than the number of spectral channels to be reconstructed, systematic reconstruction errors will occur. To investigate this effect, numerical simulations were performed to analyze the influence of the channel number on the reconstruction error.

In the simulations, the number of waveguides in the waveguide array was fixed at eight, and the maximum length difference between the waveguides was set to 420 μm. The speckle channel number was increased by varying the coefficients of Mesh A and Mesh B. Since many spectral reconstruction algorithms exist, each with different performance characteristics and parameters settings, we employ only the basic transmission-matrix method described in Eq. (7) in order to exclude the influence of the reconstruction algorithm itself and isolate the effect of the channel number on spectral reconstruction. The results are shown in Fig. S4. It can be seen that the reconstruction error gradually decreases as the speckle channel number increases.

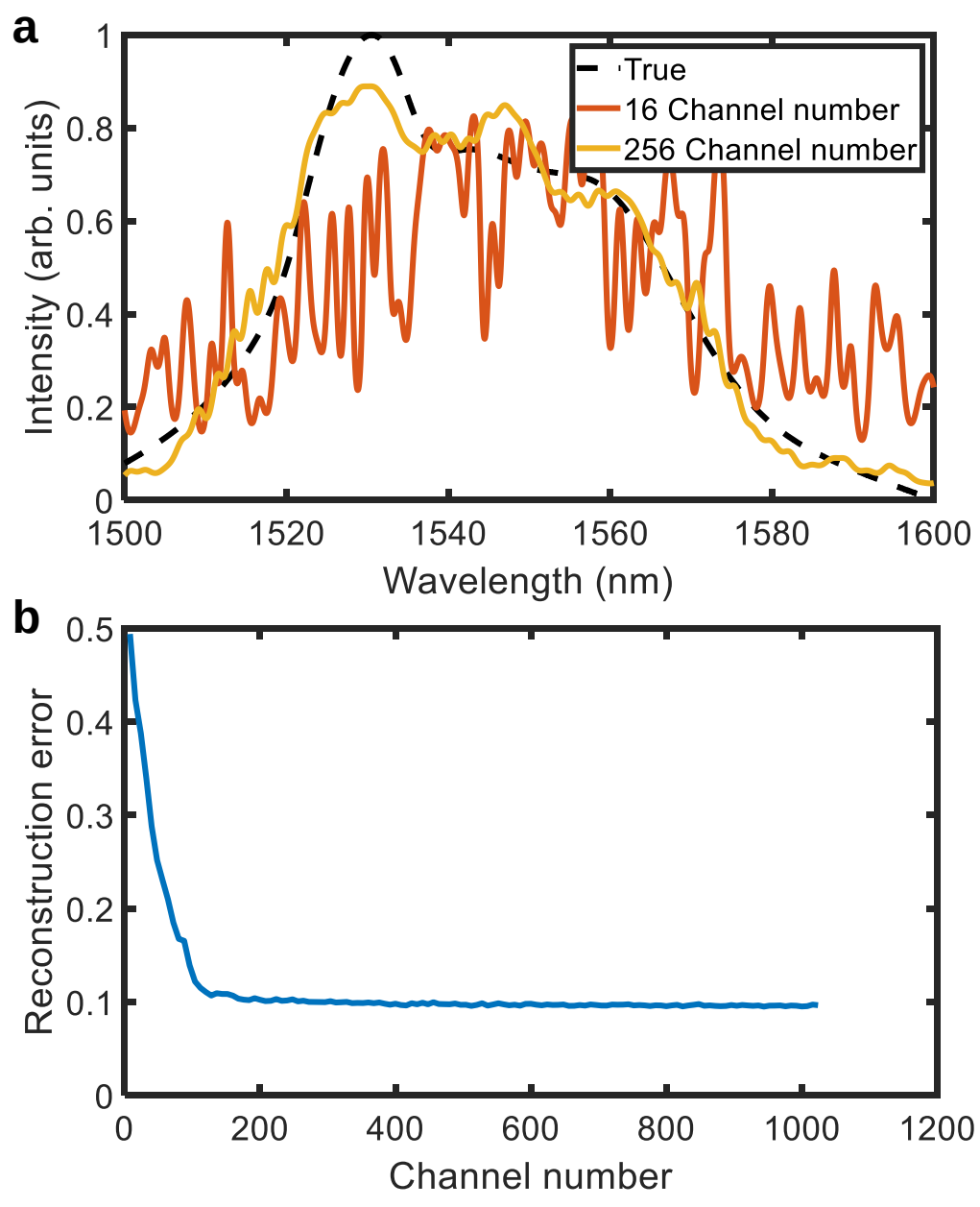

**Fig. S4. Simulation results. (a) Broadband light results of speckle reconstruction under different channel numbers. (b) Relationship between reconstruction error and speckle channel number.**

## S1.2 Theoretical analysis for spectral filtering

### S1.2.1 Principle of programmable optical filter

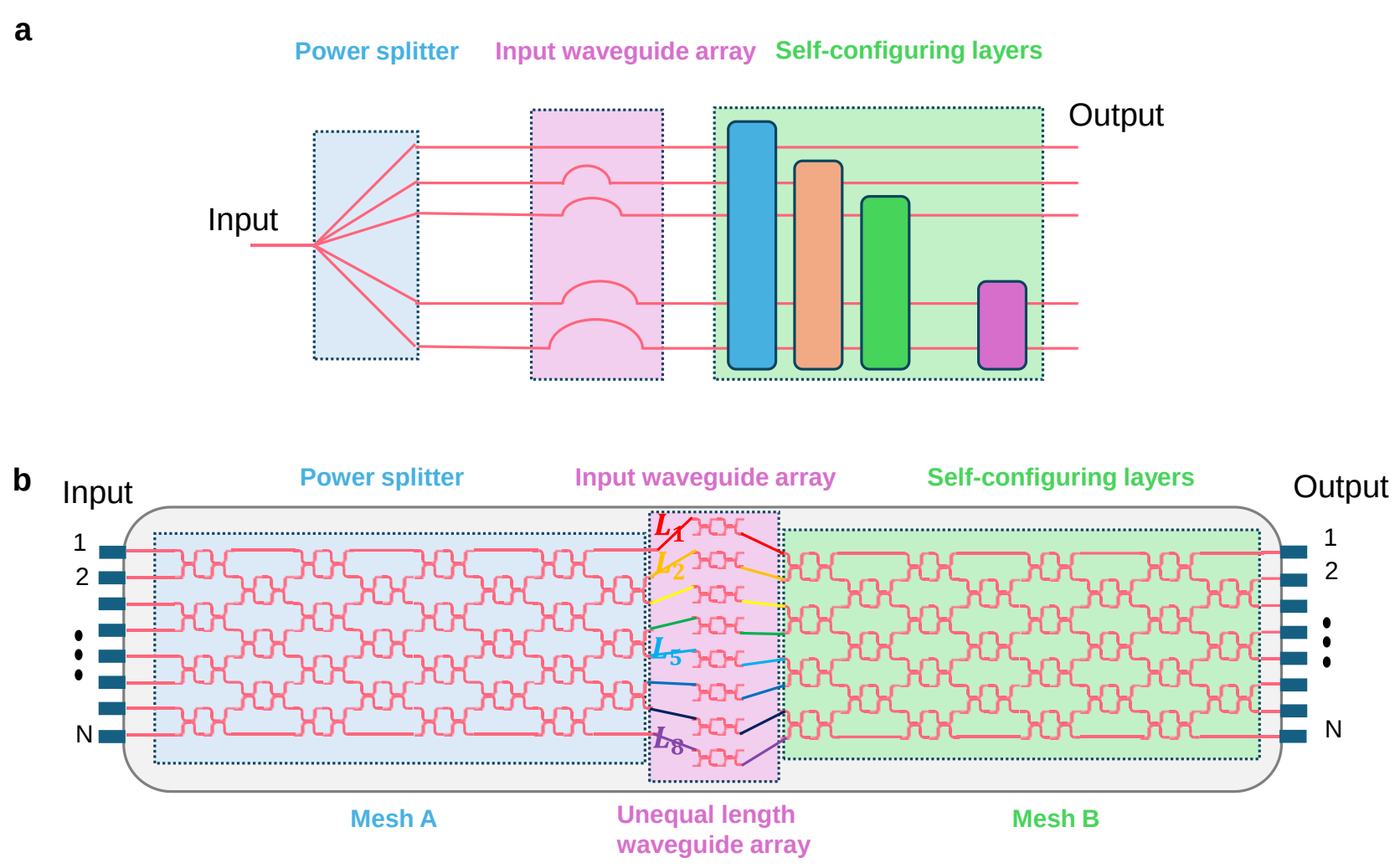

**Fig. S5. Programmable optical filter structure. (a) The structure proposed by Prof. Miller. (*Optica, 2025, 12(9): 1417-1426.*) (b) The structure proposed in this paper**

The architecture of the universal programmable and self-configuring optical filter, proposed by Prof. Miller (*Optica, 2025, 12(9): 1417-1426.*), is shown in Fig. S5(a). It consists of a power splitter, an input waveguide array, and self-configuring layers. In our implementation, Mesh A and Mesh B are used to realize the power splitter and the self-configuring layers, respectively, as shown in Fig. S5(b). Similar to the structure proposed by Prof. Miller, when the light is injected into input port $q$, the spectral response at output port $p$ of our structure can also be expressed as:

$$H_p(\delta k) = \sum_{j=1}^{m} B_{jp} A_{qj} C_j \exp\left(-i\delta k n L_j\right) \tag{8}$$

where $A_{qj}$ is the coefficient corresponding to the transmission from input port $q$ to the j-th waveguide in Mesh A, $B_{jp}$ is the corresponding coefficient in Mesh B, and $L_j$ is the length of the j-th waveguide. Eq (8) can be regarded as a "Fourier-like" series expansion, whose corresponding free spectral range and minimum achievable bandwidth are given by $f_{FSR} = 1/\delta t_0$ and $\Delta f_{min} = 1/\delta t_{max}$, respectively. $\delta t_0$ and $\delta t_{max}$ satisfy the relations $\delta t_0 = n\Delta L_{min}/c$ and $\delta t_{max} = n\Delta L_{max}/c$, respectively, where $\Delta L_{min}$ is minimum length difference in waveguide array and $\Delta L_{max}$ is maximum length difference.

As in the filter architecture proposed by Miller et al., the minimum achievable passband width is primarily determined by the maximum optical path-length difference in the waveguide array. A larger maximum path-length difference provides a larger temporal-delay span and therefore enables a narrower spectral response. For a uniformly spaced waveguide array, the spectral response is periodic, and its free spectral range (FSR) is determined by the uniform unit path-length increment. In our PSMFC, however, the eight waveguide lengths are deliberately nonuniformly spaced. Consequently, the spectral response contains multiple frequency components associated with the different pairwise path-length differences and does not exhibit a strict periodicity determined solely by the minimum length difference.

To characterize the spectral variation scale of this nonuniform array, we therefore define an effective FSR based on the average pairwise length difference. Specifically, the average length difference is calculated over all distinct pairs of waveguides as

$$\Delta L_{ave} = \frac{2}{m(m-1)} \sum_{j<k} |L_j - L_k|$$

where m=8 is the number of waveguides. For the waveguide lengths used in our device (Details in Supplementary materials S4), the maximum length difference is 423 μm, while the average pairwise length difference is 171.5 μm. The corresponding characteristic frequency scale can be estimated as

$$f_{FSR-eff} \approx \frac{c}{n_g \Delta L_{ave}}$$

where $n_g$ is the group index of the waveguide. Thus, in our nonuniform configuration, the maximum path-length difference primarily determines the minimum achievable passband width, whereas the average pairwise path-length difference is used to characterize the effective spectral repetition scale.

### S1.2.2 Filter frequency response error

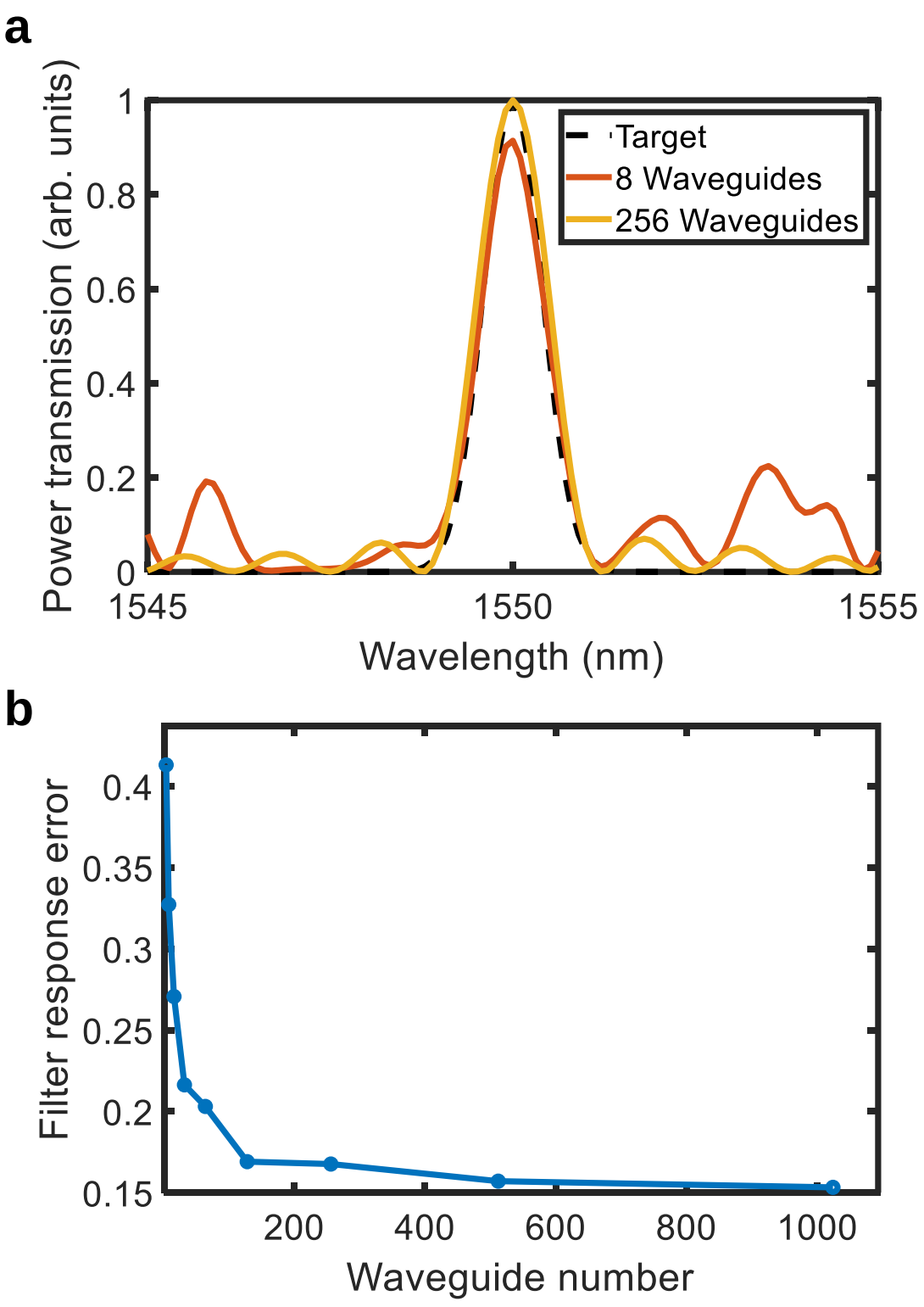

**Fig. S6. Simulation results. (a) Filter response with 8 and 256 waveguides. (b) Relationship between filter response error and waveguide number.**

From Eq. (8), it can be seen that accurately approximating an arbitrary filter frequency response requires a larger number of Fourier components. This can be achieved by using a larger-scale chip that provides more waveguides with different lengths. We performed numerical simulations to investigate the effect of the waveguide number on the accuracy of the filter response. The results are shown in Fig. S6. In the simulations, the maximum length difference between the waveguides was set to 420 μm. Under different waveguide numbers, Fig. S6(a) shows that a larger number of waveguides leads to a more refined frequency response function. To quantitatively evaluate this effect, we define the filter frequency response error as:

$$\varepsilon = \frac{\|F - F_{tar}\|_2}{\|F_{tar}\|_2} \tag{9}$$

where $F$ is the filter frequency response generated by the chip and $F_{tar}$ is the target frequency response. Fig. S6(b) shows the frequency response error for different waveguide numbers. It can be seen that increasing the number of waveguides enables a more accurate approximation of the desired filter frequency response function.

### S1.3 Influence of length distribution of the unequal length waveguide array

From the above analysis, it can be seen that in the spectrometer function, the objective is to generate random spatial speckle patterns. According to Eq. (3), non-identical length differences between pairs of waveguides lead to complex spatial speckle intensities. In contrast, for the optical filtering function, previous study (*Optica, 2025, 12(9): 1417-1426.*) have shown that a non-redundant array of waveguide lengths that increase by discrete increments can also realize the desired filter response. Therefore, in our structure, the unequal-length waveguide array does not need to follow a uniformly spaced linear increase in waveguide length.

## S2 Detail of experimental setup

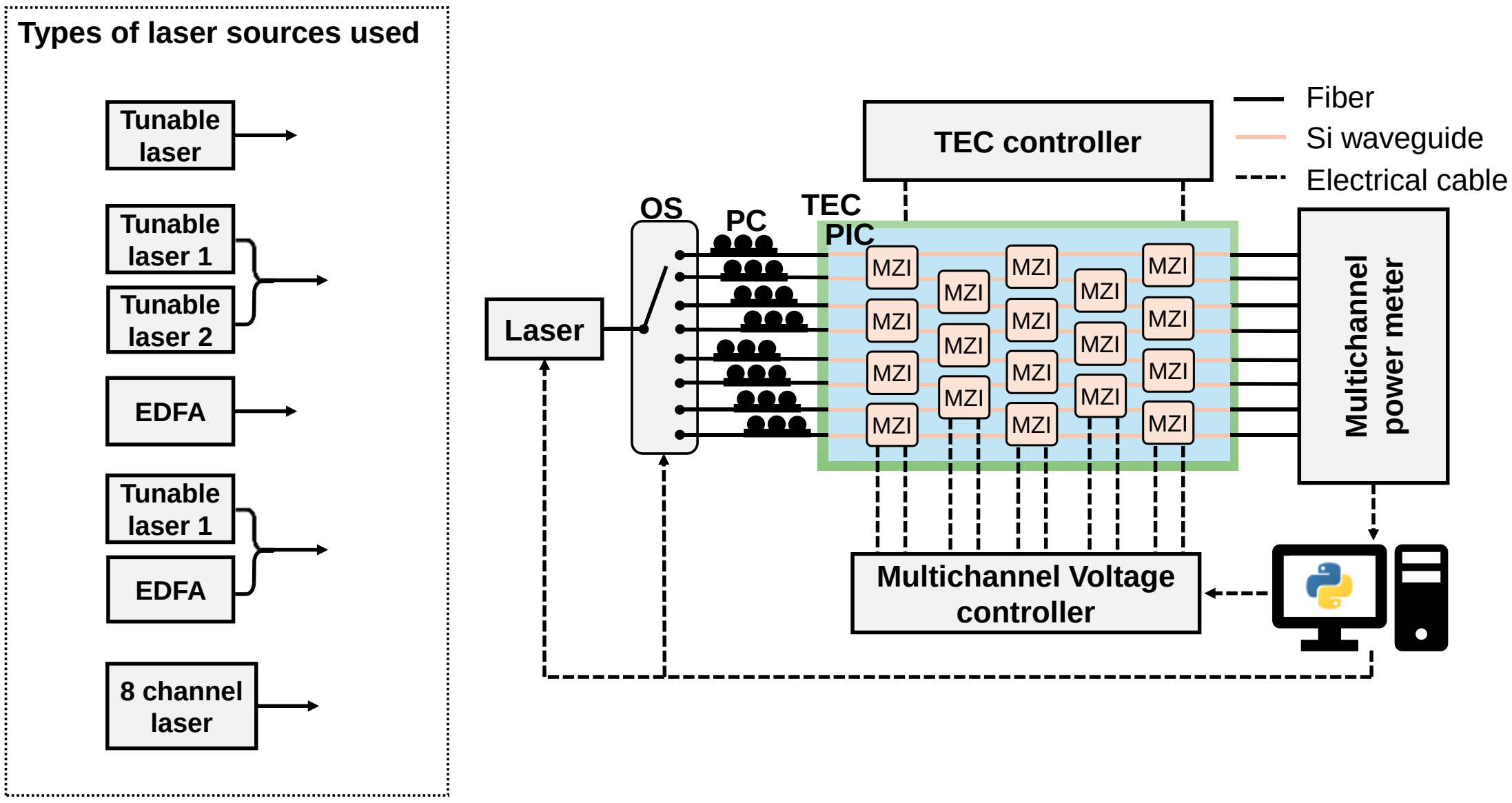

**Fig. S7. Experimental setup. OS, optical switch; PC, polarization controller; TEC,**

**Thermoelectric cooler; PIC, photonic integrated circuit; EDFA, Erbium-doped Optical Fiber Amplifier.**

## S3 Long-term stability

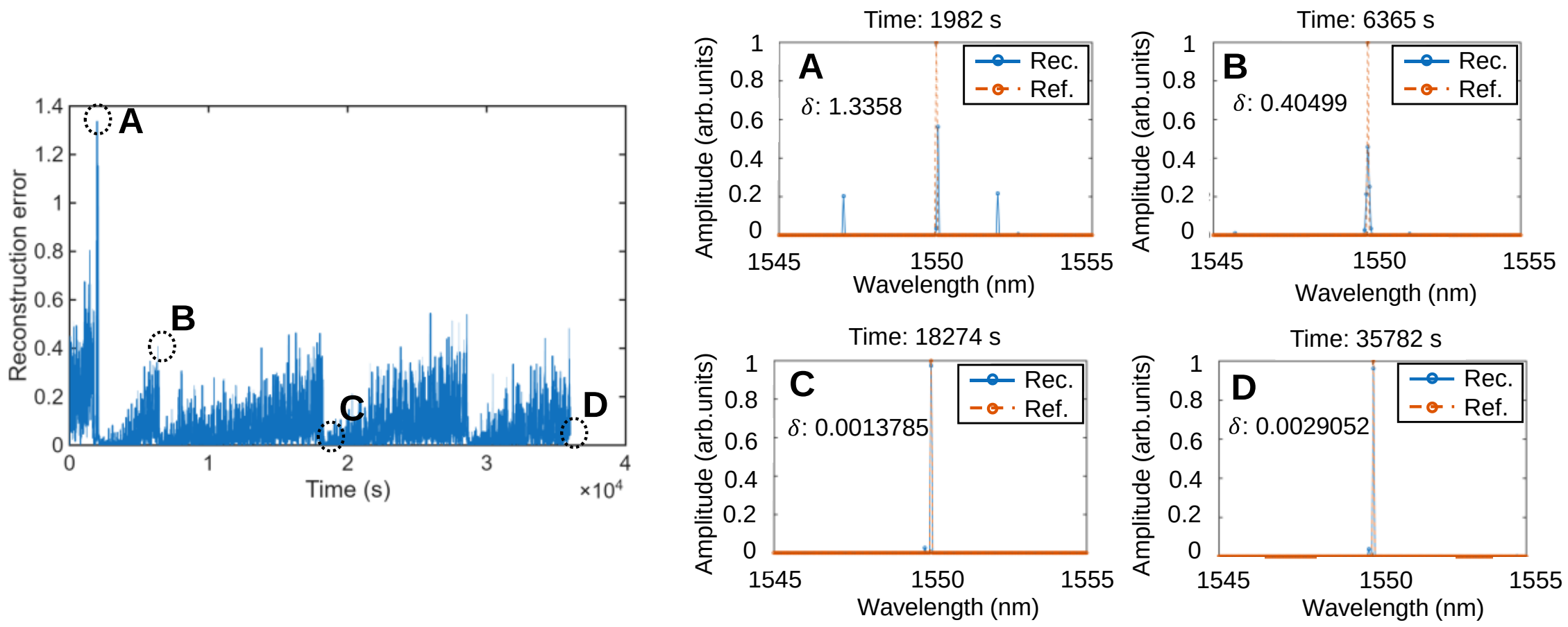

**Fig. S8. Long-term stability measurement results. Variation of spectral reconstruction error over time and reconstructed spectra at corresponding time points.**

We conducted long-term stability measurements. First, the transmission matrix of the system was established through calibration. Subsequently, a fixed-wavelength light source was used as the test input to perform continuous spectral reconstruction over an extended period. The long-term reconstruction results are shown in Fig. S8.

It can be observed that the reconstruction error exhibits a quasi-periodic variation over time, gradually increasing and then decreasing. As indicated by points A, B, C, and D in Fig. S8 the reconstruction errors at points A and B (1982 s and 6365 s) are relatively large, whereas the errors at points C and D (18274 s and 35782 s) are very small. This behavior arises from the feedback-based active temperature control system used to stabilize the chip temperature. The temperature sensor employed in the system has a resolution of 0.01℃ and an accuracy of 0.1℃. In addition, although the temperature sensor is placed close to the edge of the chip, it still takes a certain amount of time for temperature changes in the chip to propagate to the sensor. Consequently, the thermoelectric cooler (TEC) is activated only after the sensor detects a temperature change in order to restore the set temperature. As a result, the reconstruction error gradually increases until the TEC is triggered and then rapidly decreases, producing the behavior observed in Fig. S8.

The accurate spectral reconstruction at point D (around 10 hours) indicates that when the temperature remains stable, the system can maintain long-term stability. At present, the stability is mainly limited by the performance of the temperature control system. In the future, more precise temperature stabilization techniques could be employed to further improve system stability.

## S4 Detailed feature characterization of PSMFC

### S4.1 Waveguide lengths of unequal length waveguide array

For the filtering function, loss is an important performance metric. In the proposed PSMFC chip, assuming an ideal filtering response, where light at a specific wavelength is completely directed to the designated output port without any power leakage to other ports, the system loss consists of

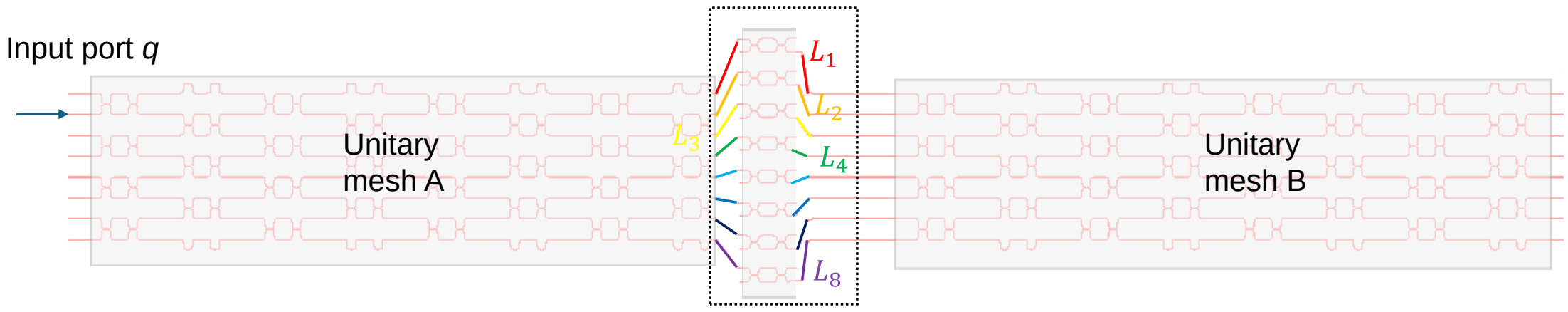

**Fig S9. Numbering diagram of an unequal length waveguide array.**

The unequal-length waveguide array consists of eight waveguides, denoted as L1–L8, arranged sequentially from top to bottom in Fig. S9. Their physical lengths are 663, 514, 393, 282, 240, 284, 382, and 508 μm, respectively. The maximum length difference among the eight waveguides is approximately 423 μm. The average length difference is approximately 171.5 μm, calculated by taking the mean of the absolute length differences over all pairwise combinations of the eight waveguides. These intentionally introduced path-length differences provide wavelength-dependent phase evolution between the two programmable unitary meshes.

### S4.2 Loss

For the filtering function, loss is an important performance metric. In the proposed PSMFC chip, assuming an ideal filtering response, where light at a specific wavelength is completely directed to the designated output port without any power leakage to other ports, the system loss consists of insertion loss and transmission loss. In our implementation, edge coupling was adopted. Experiments were conducted to measure the loss for each input port, which is defined as:

$$Loss_n = Power_{input_n} - \sum_i Power_{output_i}$$

where $Loss_n$ is the loss at the n-th input port, $Power_{input_n}$ is the optical power of incident light and $Power_{output_i}$ is the optical power at the i-th output port. The measured loss here includes only the insertion loss and transmission loss. The losses for each input port are listed in Table S1.

**Table S1. Loss at each input port**

| Input port | Loss (dB) |
|---|---|
| 1 | 8.95 ± 0.1 |
| 2 | 8.05 ± 0.1 |
| 3 | 9.25 ± 0.1 |
| 4 | 6.425 ± 0.1 |
| 5 | 7.825 ± 0.1 |
| 6 | 8.5 ± 0.1 |
| 7 | 11.7 ± 0.1 |
| 8 | 8.5 ± 0.1 |

In the current PSMFC device, typically about 80% of the optical power at the target wavelength can be directed to the designated output port. Therefore, for the filtering functionality, the overall system loss usually needs to include an additional term on top of the values listed in Table 1, corresponding to ~1 dB.

## S4.3 Voltage response of MZIs

In the PSMFC, the coefficients of Mesh A and Mesh B need to be adjusted, which requires determining the voltage response of each MZI in order to define its voltage control range. Here, we adopted a method similar to that reported in the Reference (*Journal of Lightwave Technology, 2020, 38(6): 1258-1267.*) to measure the voltage response of every MZI on the chip. The results are shown in Figs. S9–S11, corresponding to the MZIs in the unequal-length waveguide array, Mesh A, and Mesh B, respectively. It should be noted that although one faulty MZI exists in Mesh A, the performance of both the spectrometer and the optical filter is determined by the combined coefficients of Mesh A and Mesh B. Therefore, the presence of Mesh B compensates for such imperfections, ensuring that a small number of faulty MZIs in Mesh A do not significantly affect the generation of spatial speckle patterns or the frequency response of the system.

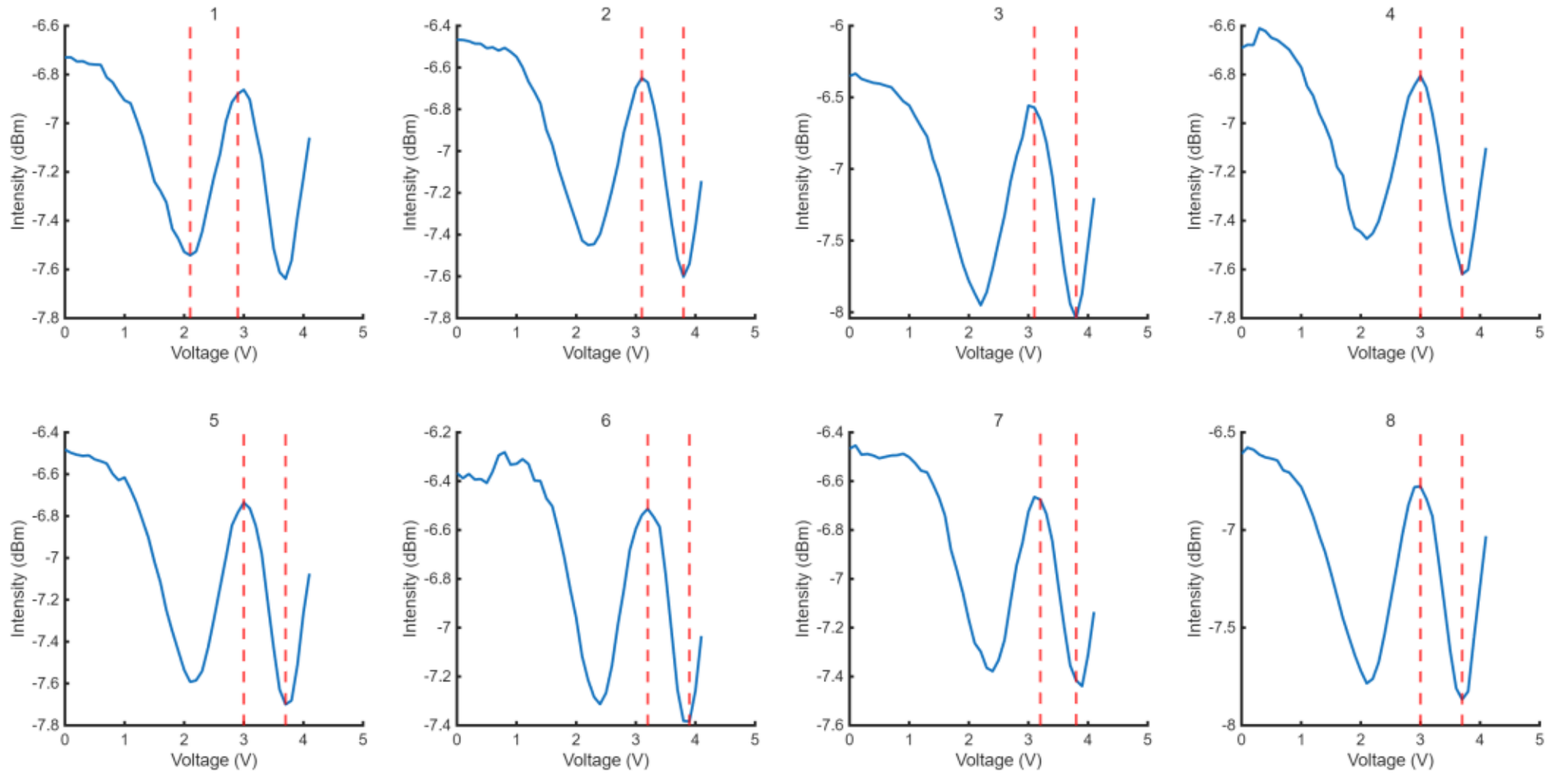

**Fig S10. Voltage response of MZIs in the unequal-length waveguide array.**

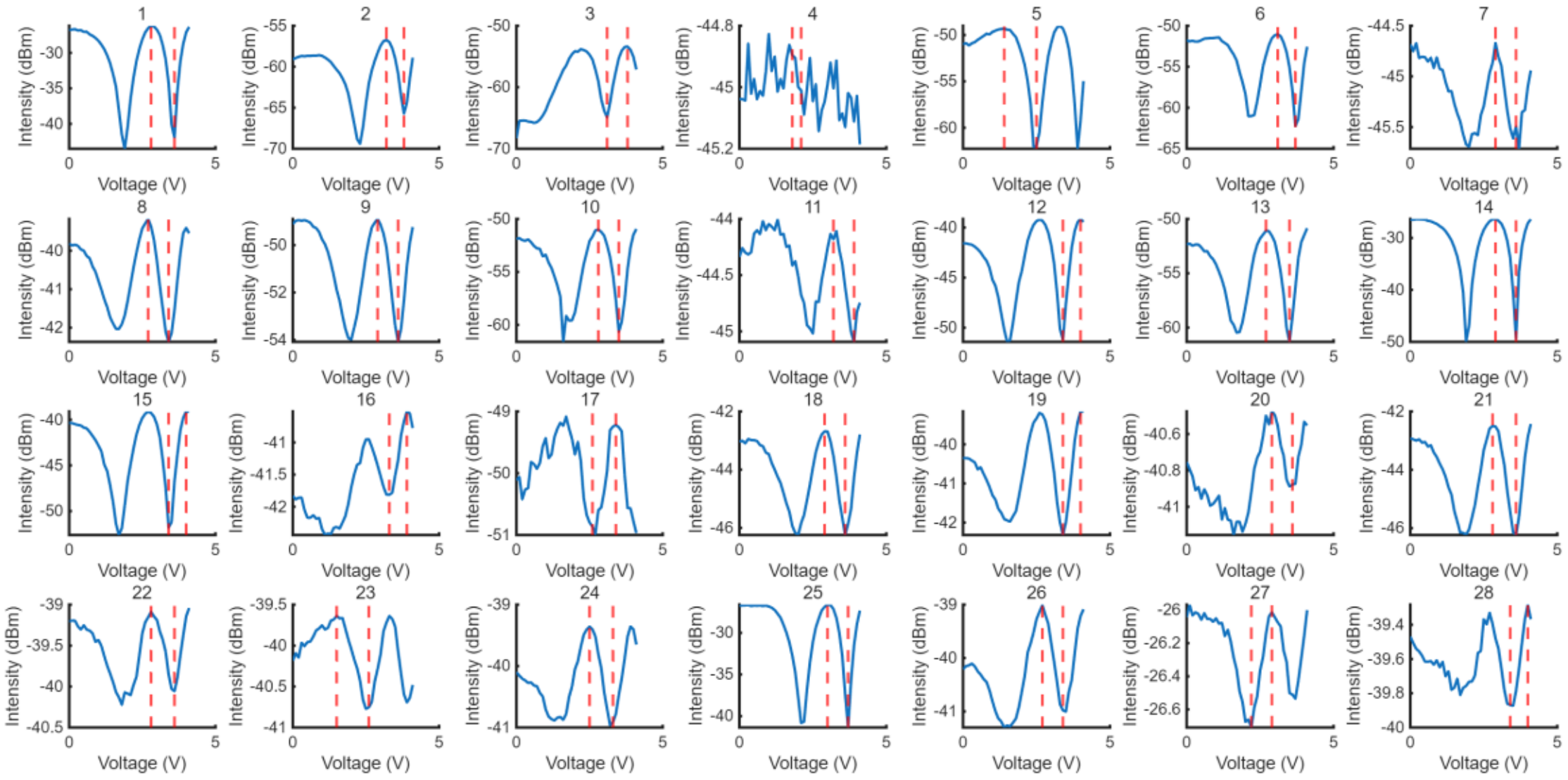

**Fig. S11. Voltage response of MZIs in the Mesh A.**

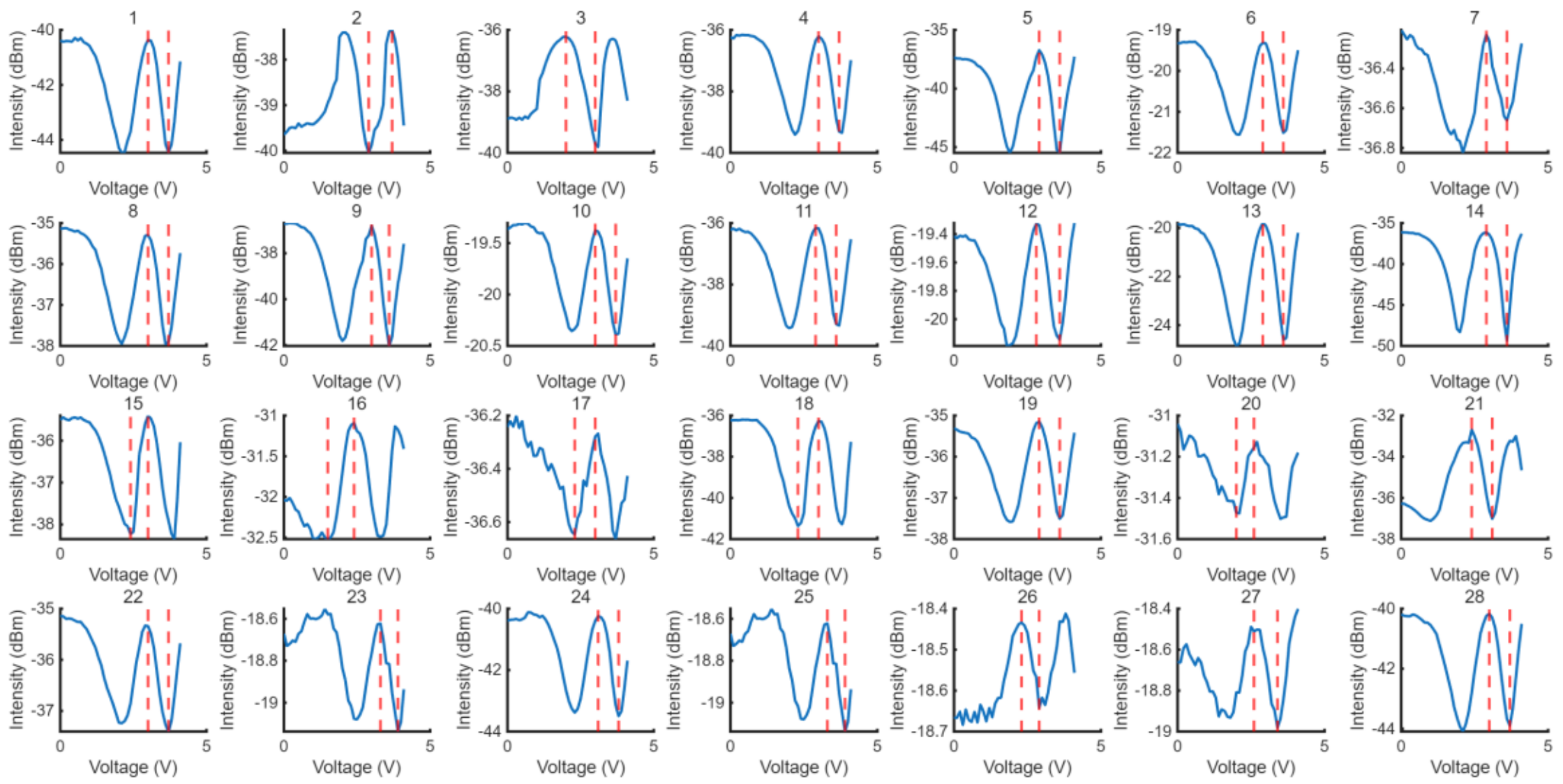

**Fig. S12. Voltage response of MZIs in the Mesh B.**